\documentclass{article}

\usepackage{arxiv}

\usepackage[utf8]{inputenc}
\usepackage[T1]{fontenc}
\usepackage{hyperref}
\usepackage{url}
\usepackage{booktabs}
\usepackage{amsmath,amssymb,amsfonts}
\usepackage{nicefrac}
\usepackage{microtype}
\usepackage{graphicx}
\usepackage[numbers]{natbib}
\usepackage{doi}
\usepackage{multirow}
\usepackage{amsthm}
\usepackage{mathrsfs}
\usepackage[title]{appendix}
\usepackage{xcolor}
\usepackage{textcomp}
\usepackage{manyfoot}
\usepackage{algorithm}
\usepackage{algorithmicx}
\usepackage{algpseudocode}
\usepackage{listings}
\usepackage{chemformula}
\usepackage{siunitx}
\usepackage[version=4]{mhchem}
\usepackage{makecell}
\usepackage{caption}
\usepackage{rotating}
\usepackage{placeins}
\usepackage{epstopdf}
\usepackage{pdflscape}
\usepackage{threeparttablex}
\usepackage{longtable}
\usepackage{array}

\DeclareSIUnit\normalcubicmetre{\text{Nm}^3}
\DeclareSIUnit\bar{\text{bar}}
\DeclareSIUnit\year{\text{y}}
\DeclareSIUnit\peryear{\text{per\,year}}
\DeclareSIUnit\perday{\text{per\,day}}
\DeclareSIUnit\gbp{\text{£}}

\renewcommand{\arraystretch}{1.15}
\newcolumntype{L}[1]{>{\raggedright\arraybackslash}p{#1}}
\newcolumntype{C}[1]{>{\centering\arraybackslash}p{#1}}

\providecommand{\backmatter}{}
\providecommand{\bmhead}[1]{\section*{#1}}
\providecommand{\parencite}[1]{\citep{#1}}
\providecommand{\bibcommenthead}{}

\title{Integrated oxy-fuel municipal solid waste incineration with flue gas recirculation, condensation filtration, and CCUS: a generic feasibility study}

\author{\parbox{0.98\textwidth}{\centering
Mikhail Kaliteevski\\
TectoHash LTD, Limassol, 100190, Cyprus\\
\texttt{mkaliteevski@gmail.com}\\[0.7em]
Leonid Chechurin\\
School of Engineering Sciences, Lappeenranta University of Technology LUT\\
Skinnarilankatu 34, Lappeenranta, 53850, Finland\\
\texttt{leonid.chechurin@lut.fi}\\[0.7em]
Maxim Permyakov\\
TectoHash LTD, Limassol, 100190, Cyprus\\
\texttt{m.permyakov@gmail.com}\\[0.7em]
Fedor Vasilyev\\
School of Engineering Sciences, Lappeenranta University of Technology LUT\\
Skinnarilankatu 34, Lappeenranta, 53850, Finland\\
\texttt{fedor.vasilyev@student.lut.fi}\\[0.7em]
Elizaveta Girshova\\
School of Engineering Sciences, Lappeenranta University of Technology LUT\\
Skinnarilankatu 34, Lappeenranta, 53850, Finland\\
\texttt{elizaveta.girshova@lut.fi}
}}
\date{}
\renewcommand{\shorttitle}{Integrated oxy-fuel municipal solid waste incineration}
\hypersetup{
pdftitle={Integrated oxy-fuel municipal solid waste incineration with flue gas recirculation, condensation filtration, and CCUS: a generic feasibility study},
pdfauthor={Mikhail Kaliteevski, Leonid Chechurin, Maxim Permyakov, Fedor Vasilyev, Elizaveta Girshova},
pdfkeywords={Municipal solid waste, Oxy-fuel combustion, Waste-to-energy, Air separation unit, CCUS}
}

\begin{document}
\maketitle

\begin{abstract}
Oxy-combustion is a promising route for improving municipal solid waste
(MSW) incineration because it removes nitrogen dilution from the oxidizing
atmosphere and produces a concentrated carbon-containing exhaust stream.
This study proposes and evaluates an integrated oxy-fuel MSW incineration
concept combining oxygen-based combustion, flue gas recirculation, heat recovery,
condensation-assisted exhaust treatment, and downstream carbon capture, utilization, or storage.   
Analysis is made for 
 \(200\,\mathrm{t\,day^{-1}}\) of as-received MSW corresponding to indicative municipal scale of about 150\,000--200\,000 people.

The process was analysed using a simplified Python
heat-balance model and Aspen Plus simulations the amount of waste. The dry combustible fraction
of MSW was represented by the empirical formula \(\ch{C_1H_{1.51334}O_{0.51837}}\), while moisture and inert ash were treated separately.

The calculations show that oxy-combustion substantially increases the attainable
combustion temperature compared with air combustion, enabling autothermal
treatment of low-calorific-value and high-moisture waste.

High combustion temperature imply utilisation of molten slag bath for combustion instead of travelling grate scheme, fly and bottom ash in this case are transformed into usable rock-like material. Flue gas recirculation is essential to shift the process from the theoretical high-temperature regime
to a practical operating range. Cooling of the oxy-fuel exhaust promotes water
condensation and produces a \ch{CO2}-rich gas suitable for further conditioning.
At the oxy-combustion dew-point temperature of approximately
\(80\,^{\circ}\mathrm{C}\), about \(2.5\,\mathrm{MW_{th}}\) of condensation heat
can be recovered from the oxy-fuel flue gas.
For the reference \(1350\,^{\circ}\mathrm{C}\) FGR case,  the gross steam-cycle electric
output is approximately \(3.7\,\mathrm{MW}\). 

Further, high supersaturation of water vapour in flue gas ensures that each particle of dust serves a centre of condensation of water. High concentration of dust particles, typically ${10^{13} /m^3}$ ,   
ensures condensation of water on the dust particle at the time scale of the seconds, thus providing capture of dust particles. 
Volume normalized surface area of the water drops is of the order $500 m^{-1}$ and provide capture of \ch{SO2}, \ch{SO3}, \ch{HCl}, \ch{HF} and \ch{NO2}.
Utilization of the recovered
condensation heat for process preheating is assumed to provide an additional
electric-power contribution of approximately \(0.75\,\mathrm{MW_e}\),
corresponding to \(30\%\) of the recovered thermal power. The resulting total
electric output is approximately \(4.5\,\mathrm{MW}\). After internal auxiliary
loads, ASU oxygen production at \(0.265\,\mathrm{kWh\,kg^{-1}_{O_2}}\), and an
estimated CCUS penalty, the remaining net electric power is approximately
\(1.1\,\mathrm{MW}\). The results indicate that the proposed integrated
configuration can support oxygen supply, downstream carbon management, and heat
recovery, although its feasibility remains strongly sensitive to ASU performance
and plant scale.
\end{abstract}


\section{Introduction}\label{sec1}

Waste accumulation is a growing problem. As global population and consumption both increase, waste generation has reached level of more than 2 billion tonnes now \cite{ref:1} and is expected to rise by \SI{70}{\percent} by 2050 \cite{ref:2}. This excessive waste is having a detrimental impact on the environment \cite{ref:1}. The problem of waste becomes an issue that is likely to dominate the world agenda in the coming century, and there are three major ways to deal with: sorting, reusing and recycling; waste-to-energy; landfilling.

While sorting and recycling are preferable, their expansion requires additional
sorting facilities and workforce and may also introduce occupational health risks
\cite{ref:3}. Landfilling is the most hazardous yet unfortunately common waste management method, having growing footprint and triggering uncontrolled chemical and biological processes that cause self-heating and ignition.     Garbage burning in landfills can release particulate matter and toxic organic
pollutants. Dioxins and furans are among the environmentally important products
of uncontrolled waste combustion \cite{ref:5}. 

Chemical or biological transformation of garbage into fuel or soil can become a good decision, but significant diversity of chemical and morphological varieties of waste makes it unlikely to develop a technology that provides transformation of plastic, paper, food, and other types of waste by one chemical or biological process.

Waste-to-energy technologies involve processes such as incineration, gasification, and pyrolysis \cite{ref:7}. These methods can face several challenges, including the high costs \cite{ref:8}, as well as environmental concerns \cite{ref:9}.

At present, waste-to-energy technologies are usually adjusted to a specific type of waste, so some kinds of waste are excluded. Improving waste incineration includes expanding the range of acceptable waste,
reducing hazardous solid residues, and producing a more concentrated
\ch{CO2}-containing exhaust stream that can be further treated for carbon capture
\cite{fu2025conceptual}.

Municipal solid waste incineration can produce hazardous pollutants, including
dioxins, which require careful control of combustion conditions and flue-gas
treatment \cite{ref:39}. Most conventional MSW incineration plants use a travelling-grate combustion
chamber supplied with atmospheric air \cite{ref:14}. This type of furnace is
widely used for continuous combustion of heterogeneous municipal waste
\cite{ref:15}. The released heat is normally recovered through steam production
and subsequent power generation.

Flue-gas treatment is usually performed in several stages. Dust and particulate
matter are first removed by mechanical or filtration systems \cite{ref:16}.
Acidic and sulphur-containing compounds require additional gas-cleaning steps
\cite{ref:17}. Adsorption and other treatment methods can also be used for
removal of organic and trace contaminants \cite{ref:18}. Nitrogen oxides require
a separate removal step, for example by chemical absorption or related
technologies \cite{ref:20}. Dioxin formation remains an important concern in MSW incineration and requires
careful control of combustion conditions and flue-gas treatment \cite{ref:39}. Further, incineration plants emit carbon dioxide, contributing to global warming.

In this article, we evaluate an integrated oxy-fuel MSW incineration concept with flue gas recirculation, condensation-assisted exhaust treatment, heat recovery, oxygen production, and downstream CCUS at a medium municipal scale, approximately corresponding to the residual waste generated by \(150{,}000\text{--}200{,}000\) people.

\section{Literature and industrial landscape}\label{sec2}

Different approaches to oxygen-assisted waste combustion have been studied.
Oxygen-enhanced waste incineration has been analysed from both economic and
environmental points of view \cite{ref:24}. Integrated waste-to-energy systems
combining oxygen enrichment with an air separation unit have also been proposed
\cite{ref:26}. Experimental studies have examined how increasing the oxygen
content of the oxidizer affects combustion characteristics \cite{ref:27}.

Experimental work has also been performed at different scales. Kumar and Singh
studied oxygen-enriched co-combustion in a 20 kW bubbling fluidized-bed combustor
\cite{ref:28}. Oxygen-blown conversion of combustible waste has also been studied
in a fixed-bed gasifier \cite{ref:29}. At industrial scale, Ma et al. investigated
oxygen-enriched combustion of municipal solid waste in an 8 MW$_{\mathrm{th}}$
waste-to-energy plant \cite{ref:23}. A broader review of oxygen-fuel combustion
technologies is provided in \cite{ref:30}.

Increasing oxygen content can raise the combustion temperature, reduce unburned
carbon, and affect pollutant formation. In a pilot-scale study of sewage sludge
co-incineration with municipal solid waste, oxygen enrichment improved combustion
performance and reduced HCl and SO$_2$ emissions \cite{ref:22}.

Recent studies show that oxy-fuel MSW incineration should be evaluated as an integrated system rather than only at the furnace level. Fu et al. presented a conceptual design of an MSW WtE plant with oxy-combustion \ch{CO2} capture and reported the possibility of negative emissions of 510.6 kg \ch{CO2} per ton of MSW, depending on waste composition and process configuration \cite{fu2025conceptual}. Their analysis also showed that air leakage, oxygen purity, and residual oxygen at the furnace outlet strongly affect impurity levels in the \ch{CO2}-rich flue gas and the energy demand of downstream purification. Therefore, the system boundary should include oxygen supply, flue gas recycling, \ch{CO2} purification, and auxiliary energy use.

Another emerging direction is coupling oxygen-fuel MSW combustion with renewable electricity and water electrolysis. Huang, Jiang and Tong proposed an oxygen-fuel combustion system driven by electrolytic oxygen production from renewable energy, framing oxygen as a co-product of renewable hydrogen production rather than only as an industrial input from air separation \cite{huang2024design}. Re et al. further studied co-location of oxy-fuel waste combustion with electrolyzers, considering a 229 MWth oxy-combustion waste CHP plant integrated with wind power, cryogenic air separation, alkaline electrolysis, hydrogen and oxygen storage, and grid expansion \cite{re2025co}. They concluded that such co-location could support green hydrogen market penetration and estimated that scaling one proposed configuration to existing Finnish WtE plants could cover up to 27\% of Finland’s green hydrogen target for 2030.

\begin{figure}[htb]
 \centering
 \includegraphics[width=0.95\linewidth]{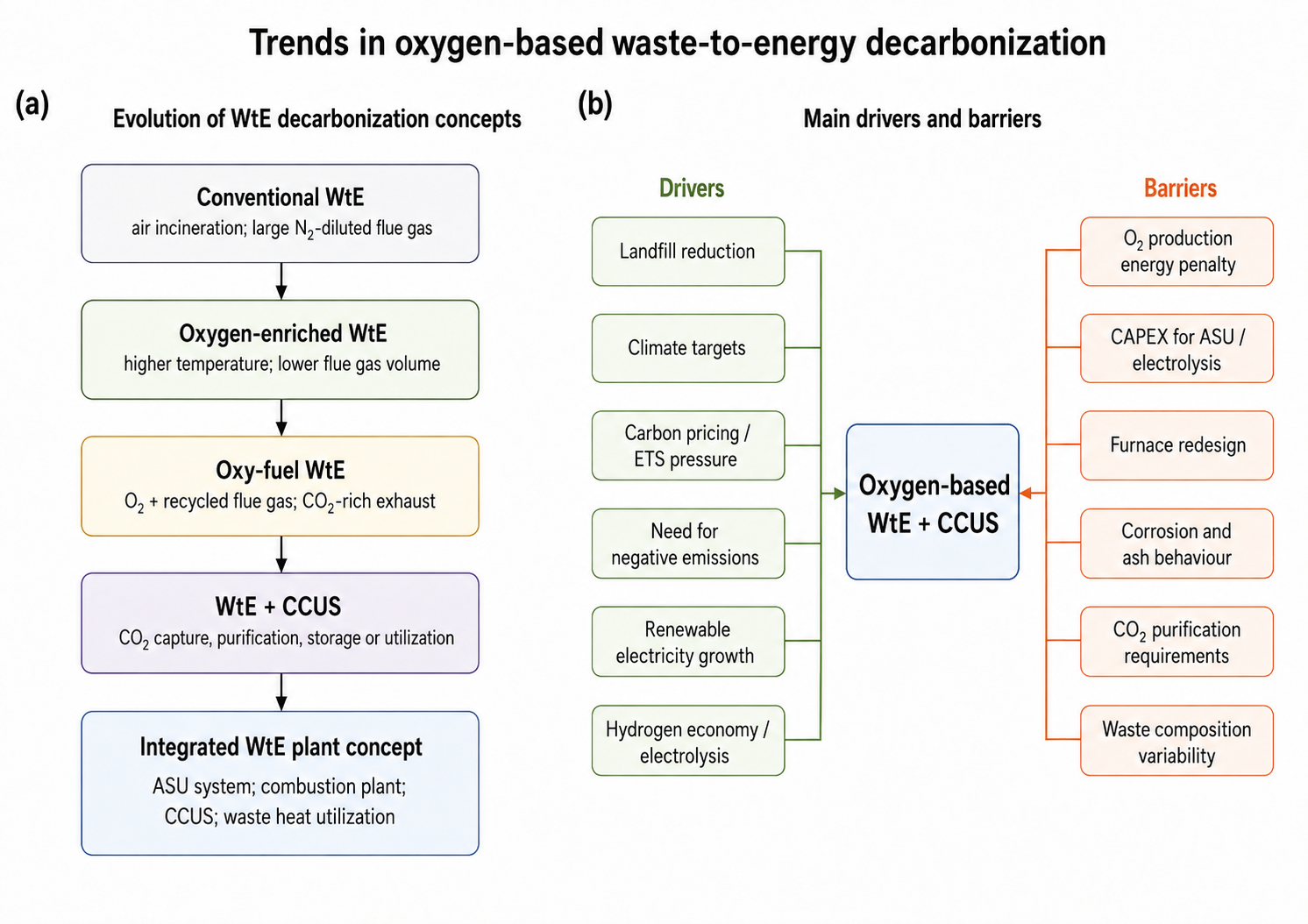}
 \caption{Trends in oxygen-based waste-to-energy decarbonization. (a) Evolution of waste-to-energy concepts from conventional air incineration with nitrogen-diluted flue gas to oxygen-enriched combustion, oxy-fuel combustion, WtE with CCUS, and integrated plant concepts combining an air separation unit, combustion plant, \ch{CO2} capture and utilization/storage, and waste heat utilization. (b) Main drivers and barriers for oxygen-based WtE with CCUS, including landfill reduction, climate targets, carbon pricing, negative-emission potential, renewable electricity growth, and hydrogen/electrolysis integration, as well as oxygen production penalties, ASU or electrolysis CAPEX, furnace redesign, corrosion and ash behavior, \ch{CO2} purification requirements, and waste composition variability.}
 \label{fig:literature_landscape}
\end{figure}

Figure~\ref{fig:literature_landscape} summarizes the conceptual evolution of oxygen-based waste-to-energy systems and the main drivers and barriers for integrated WtE--CCUS plants.

These studies show a shift from isolated combustion analysis toward integrated system design. Earlier oxy-fuel research mainly addressed coal-fired power plants, boiler retrofits, flame-temperature control, oxygen production penalties, and \ch{CO2} purification. More recent work extends these concepts to heterogeneous waste fuels, where additional challenges arise from variable moisture content, ash composition, chlorine, heavy metals, and fluctuating calorific value. Thus, oxy-fuel combustion of MSW is not a direct transfer of coal-combustion knowledge, but requires specific attention to waste composition, corrosion, slagging, flue gas cleaning, and ash management.

Oxy-fuel combustion is also increasingly discussed within broader carbon-management frameworks, including biomass combustion, BECCS, chemical looping combustion, gasification, and carbon capture integration \cite{vasileiadou2025advancements}. This framing is important because MSW contains both fossil-derived and biogenic carbon. Capturing \ch{CO2} from WtE plants can therefore reduce fossil emissions and may potentially generate negative emissions from the biogenic fraction of the waste stream.

From an industrial perspective, most current WtE decarbonization projects are based on post-combustion \ch{CO2} capture rather than full oxy-fuel combustion, because post-combustion capture can be retrofitted to existing air-fired plants, whereas oxy-fuel combustion requires deeper modification of the furnace, oxidizer supply, flue gas recycling, heat-transfer design, and control system. Nevertheless, existing WtE--CCUS projects are highly relevant because they demonstrate the emerging \ch{CO2} transport, storage, utilization, and regulatory landscape into which oxy-fuel systems would need to fit.

Important industrial examples include the Klemetsrud WtE plant in Oslo, operated by Hafslund Celsio, where a full-scale carbon capture facility involving Aker Solutions and SLB Capturi is expected to become operational in 2029 and to capture around 90\% of the flue-gas \ch{CO2} \cite{HafslundCelsio2025KlemetsrudCCS}. In the Netherlands, Twence captures about 100,000 tonnes of \ch{CO2} per year for utilization in building materials and bottom ash treatment, demonstrating a local circular-economy business model \cite{SLBCapturi2026TwenceCCU}. Amager Bakke in Copenhagen is another relevant case. ARC aims to make the plant
\ch{CO2}-neutral through carbon capture and has tested the technology in smaller
demonstration units \cite{ARC2026AmagerBakkeCarbonCapture}. In 2025, ARC started
a collaboration with E.ON on the CopenCapture project for large-scale
\ch{CO2} capture at Amager Bakke \cite{EON2025CopenCapture}.

\FloatBarrier
\section{Methodology}\label{sec3}
This study combines literature analysis, thermodynamic modelling, Aspen Plus process simulation, and feasibility-level energy and cost estimates. Recent academic and industrial developments in oxygen-enriched and oxy-fuel waste-to-energy systems, CCUS integration, and oxygen-production technologies were reviewed to define the main process functions: oxygen-based combustion, flue gas recirculation, heat recovery, condensation-based exhaust treatment, and downstream \(\mathrm{CO_2}\) conditioning.

The thermodynamic behaviour of MSW combustion was analysed using two complementary approaches. First, a simplified Python model calculated combustion temperatures from an integral heat-capacity balance. The MSW feed was represented as dry combustible matter, moisture, and inert ash. The dry combustible fraction was described by the Aspen-derived empirical formula \(\mathrm{C_1H_{1.51334}O_{0.51837}}\). Its lower heating value was estimated using a Dulong-type correlation, and the lower heating value of as-received MSW was obtained by normalizing dry-fraction energy content by the total waste mass flow. Temperature-dependent heat capacities were used for gaseous products, while ash was treated as inert \(\mathrm{SiO_2}\) with constant heat capacity.

Second, Aspen Plus was used as an independent process-simulation reference. MSW was represented as separate dry combustible, moisture, and ash streams mixed before entering an oxygen-fed combustion chamber. Complete oxidation was assumed. The product stream was used to evaluate reference temperatures, composition, oxygen demand, heat recovery, and power balance. Aspen results were compared with Python calculations at different flue gas recirculation rates to verify the main thermodynamic trends.

The exhaust-gas condensation analysis used water thermophysical properties from CoolProp. Condensation efficiency was calculated from the difference between water-vapour partial pressure and saturation pressure at the selected cooling temperature. This was used to compare air- and oxy-combustion exhaust streams and evaluate condensation-assisted removal of water and soluble or particle-associated impurities.

The feasibility-level energy balance was based on the Aspen Plus reference case with flue gas recirculation and a target combustion temperature of \(1350\,^{\circ}\mathrm{C}\). Gross electric output was calculated from the steam-turbine stages, while internal auxiliary loads included oxygen compression, flue gas recirculation compression, and pump work. Additional parasitic loads for oxygen production and downstream \(\mathrm{CO_2}\) conditioning/CCUS were estimated from the fresh oxygen and \(\mathrm{CO_2}\) mass flow rates using assumed specific electricity demands.

Finally, ASU scale effects were evaluated using reported commercial, literature, benchmark, and prospective oxygen-production data. These data were compared with the oxygen demand of the reference oxy-fuel MSW case. ASU capital cost was estimated using a scaling relation based on installed oxygen-production capacity, and the capacity-normalized ASU cost was calculated by dividing total ASU cost by installed daily oxygen-production capacity.

\FloatBarrier

\section{Thermodynamic aspects of oxygen-based MSW combustion}\label{sec4}
The thermal behaviour of municipal solid waste combustion was analysed using two complementary approaches: a simplified Python model based on an integral heat-capacity balance and Aspen Plus process simulations. The purpose of this comparison was to evaluate whether the simplified model reproduces the main temperature trends obtained from the process simulator and can therefore be used for parametric analysis of oxygen-based waste combustion.

The fuel energy input was expressed using the lower heating value of the total as-received MSW stream, \(LHV_{\mathrm{MSW}}\). This definition includes the effect of the actual moisture and ash content of the waste stream and is therefore suitable for comparing different waste qualities on the same basis. The total MSW stream was treated as a mixture of dry combustible matter, moisture, and inert ash. A qualitative summary of the ignition behaviour, limiting oxygen index, lower heating value, and expected oxygen-enrichment response of representative MSW components is provided in Appendix A, Supplementary Table S1.

The dry combustible fraction of MSW was represented by the empirical formula taken from the Aspen Plus library:
\begin{equation}
\mathrm{C_1H_{1.51334}O_{0.51837}} .
\end{equation}

For a general empirical fuel formula
\begin{equation}
\mathrm{C_aH_bO_c},
\end{equation}
the stoichiometric oxygen requirement is
\begin{equation}
\nu_{\mathrm{O_2,st}} =
a + \frac{b}{4} - \frac{c}{2}.
\end{equation}

For the Aspen-derived dry MSW formula, this gives
\begin{equation}
\nu_{\mathrm{O_2,st}}
=
1+\frac{1.51334}{4}-\frac{0.51837}{2}
\approx 1.119
\ \mathrm{mol\ O_2\,mol^{-1}\ fuel}.
\end{equation}

Accordingly, the stoichiometric oxidation of the dry combustible fraction can be written as
\begin{equation}
\mathrm{C_1H_{1.51334}O_{0.51837}}
+
1.119\,\mathrm{O_2}
\rightarrow
\mathrm{CO_2}
+
0.7567\,\mathrm{H_2O}.
\end{equation}

The oxidizer supply was described using the stoichiometric ratio
\begin{equation}
SR =
\frac{n_{\mathrm{O_2,supplied}}}
{n_{\mathrm{O_2,stoichiometric}}},
\end{equation}
where \(SR = 1\) corresponds to stoichiometric combustion and \(SR > 1\) corresponds to excess oxygen supply. In the oxy-combustion case, the product mixture was assumed to contain \(\mathrm{CO_2}\), \(\mathrm{H_2O}\), excess \(\mathrm{O_2}\), and inert ash represented as \(\mathrm{SiO_2}\). In the air-combustion case, nitrogen introduced with air was included as
\begin{equation}
n_{\mathrm{N_2}} =
3.76\,SR\,n_{\mathrm{O_2,stoichiometric}}.
\end{equation}

The lower heating value of the dry combustible fraction was estimated from the empirical formula using a Dulong-type elemental correlation. First, the molar mass of the dry combustible fraction\citep{CIAAWAtomicWeights}:
\begin{equation}
M_{\mathrm{fuel}}
=
12.011
+
1.51334 \times 1.008
+
0.51837 \times 15.999
\approx
21.83\ \mathrm{g\,mol^{-1}}.
\end{equation}
The corresponding elemental mass fractions were then obtained as
\begin{equation}
w_{\mathrm{C}}
=
\frac{12.011}{M_{\mathrm{fuel}}},
\qquad
w_{\mathrm{H}}
=
\frac{1.51334 \times 1.008}{M_{\mathrm{fuel}}},
\qquad
w_{\mathrm{O}}
=
\frac{0.51837 \times 15.999}{M_{\mathrm{fuel}}}.
\end{equation}
For \(\mathrm{C_1H_{1.51334}O_{0.51837}}\), this gives approximately
\begin{equation}
w_{\mathrm{C}} \approx 0.550,
\qquad
w_{\mathrm{H}} \approx 0.070,
\qquad
w_{\mathrm{O}} \approx 0.380.
\end{equation}
The higher heating value of the dry combustible fraction was estimated using the Dulong-type correlation:
\begin{equation}
HHV_{\mathrm{dry}}
=
33.87w_{\mathrm{C}}
+
144.2
\left(
w_{\mathrm{H}}-\frac{w_{\mathrm{O}}}{8}
\right),
\end{equation}
where \(HHV_{\mathrm{dry}}\) is expressed in \(\mathrm{MJ\,kg^{-1}}\). The lower heating value was then calculated by subtracting the latent heat associated with water formed from hydrogen:
\begin{equation}
LHV_{\mathrm{dry}}
=
HHV_{\mathrm{dry}}
-
2.442 \times 9w_{\mathrm{H}}.
\end{equation}
For the empirical formula \(\mathrm{C_1H_{1.51334}O_{0.51837}}\), this gives
\begin{equation}
LHV_{\mathrm{dry}}
\approx
20.26\ \mathrm{MJ\,kg^{-1}}.
\end{equation}

The combustion temperature was calculated from an integral heat-capacity balance. For the dry combustible fraction, the released heat was calculated from its lower heating value, while the initial moisture content was included through the heat required for water evaporation. Thus, the net heat available for heating the combustion products was written as
\begin{equation}
Q_{\mathrm{net}}
=
\gamma
\left(
m_{\mathrm{dry}}LHV_{\mathrm{dry}}
-
m_{\mathrm{H_2O}}L_{\mathrm{vap}}
\right),
\end{equation}
where \(m_{\mathrm{dry}}\) is the mass of dry combustible matter, \(LHV_{\mathrm{dry}}\) is the lower heating value of the dry combustible fraction, \(m_{\mathrm{H_2O}}\) is the mass of moisture in the as-received waste, \(L_{\mathrm{vap}}\) is the latent heat of water evaporation, and \(\gamma\) is the heat-retention factor of the combustion chamber.

The combustion temperature was then obtained from
\begin{equation}
\gamma
\left(
m_{\mathrm{dry}}LHV_{\mathrm{dry}}
-
m_{\mathrm{H_2O}}L_{\mathrm{vap}}
\right)
=
\sum_j n_{j,\mathrm{out}}
\int_{T_0}^{T_{\mathrm{final}}}
c_{p,j}(T)\,dT
+
m_{\mathrm{ash}}c_{\mathrm{ash}}
\left(
T_{\mathrm{final}}-T_0
\right).
\end{equation}
Here, \(T_{\mathrm{final}}\) is the calculated combustion temperature, \(T_0\) is the reference temperature, \(n_{j,\mathrm{out}}\) is the molar amount of product species \(j\), and \(c_{p,j}(T)\) is its temperature-dependent molar heat capacity. The product species include \(\mathrm{CO_2}\), \(\mathrm{H_2O}\), and excess \(\mathrm{O_2}\) for oxy-combustion, with additional \(\mathrm{N_2}\) for air combustion. The ash fraction was represented as inert \(\mathrm{SiO_2}\). Its heat capacity was approximated as constant, so the ash sensible heat term was written as \(m_{\mathrm{ash}}c_{\mathrm{ash}}(T_{\mathrm{final}}-T_0)\).

For positioning the Aspen reference point on the \(LHV_{\mathrm{MSW}}\)--\(SR\) temperature maps, the lower heating value of the total as-received MSW stream was calculated by normalizing the energy content of the dry combustible fraction by the total MSW mass flow:
\begin{equation}
LHV_{\mathrm{MSW}}
=
\frac{
\dot m_{\mathrm{dry}} LHV_{\mathrm{dry}}
}
{
\dot m_{\mathrm{dry}}+\dot m_{\mathrm{H_2O}}+\dot m_{\mathrm{ash}}
}.
\end{equation}
Here, \(\dot m_{\mathrm{dry}}\), \(\dot m_{\mathrm{H_2O}}\), and \(\dot m_{\mathrm{ash}}\) are the mass flow rates of the dry combustible fraction, moisture, and ash, respectively. This definition gives the lower heating value per unit mass of the total as-received waste feed. The heat required for evaporation of the initial moisture was accounted for separately in the heat-balance equation through the term \(m_{\mathrm{H_2O}}L_{\mathrm{,vap}}\).

The Aspen reference case was defined by the following feed composition and total as-received MSW flow rate. This total flow corresponds to approximately \(200~\mathrm{t\,day^{-1}}\) of waste. Assuming an illustrative municipal waste generation rate of \(1~\mathrm{kg\,person^{-1}\,day^{-1}}\), this scale is equivalent to the residual waste generated by about \(2.0\times10^5\) people, before accounting for local sorting, reuse, and recycling rates:
\begin{equation}
\begin{aligned}
\dot m_{\mathrm{dry}} &= 4186~\mathrm{kg\,h^{-1}},\\
\dot m_{\mathrm{H_2O}} &= 2070~\mathrm{kg\,h^{-1}},\\
\dot m_{\mathrm{ash}} &= 2020~\mathrm{kg\,h^{-1}},\\
\dot m_{\mathrm{MSW}} 
&= \dot m_{\mathrm{dry}}+\dot m_{\mathrm{H_2O}}+\dot m_{\mathrm{ash}} = 8276~\mathrm{kg\,h^{-1}}
\approx 200~\mathrm{t\,day^{-1}}.
\end{aligned}
\label{eq:aspen_reference_feed}
\end{equation}
Using \(LHV_{\mathrm{dry}} = 20.26\ \mathrm{MJ\,kg^{-1}}\), this gives
\begin{equation}
LHV_{\mathrm{MSW}}
=
\frac{
4186 \times 20.26
}
{
4186+2070+2020
}
\approx
10.25\ \mathrm{MJ\,kg^{-1}}.
\end{equation}

The same Aspen reference case was used to define the marked point in the ash--moisture maps:
\begin{equation}
\frac{\dot m_{\mathrm{ash}}}{\dot m_{\mathrm{dry}}}
=
\frac{2020}{4186}
\approx 0.483,
\qquad
\frac{\dot m_{\mathrm{H_2O}}}{\dot m_{\mathrm{dry}}}
=
\frac{2070}{4186}
\approx 0.495.
\end{equation}

The value of \(T_{\mathrm{final}}\) was found numerically by solving the heat-balance equation for each selected combination of \(LHV_{\mathrm{MSW}}\), stoichiometric ratio, moisture content, ash content, and flue gas recirculation rate.

The Aspen Plus reference model was implemented as a simplified steady-state combustion scheme. The MSW feed was represented by three separate streams: dry combustible matter, moisture, and inert ash. The dry combustible fraction was described by the empirical formula \(\mathrm{C_1H_{1.51334}O_{0.51837}}\), the moisture fraction was introduced as water, and the ash fraction was represented as inert \(\mathrm{SiO_2}\). These streams were combined in a feed-mixing block and supplied to the combustion chamber together with a separate oxygen stream.

The oxygen stream was passed through a compression and heat-recovery line before entering the combustion chamber, allowing the effect of oxidizer preheating and process integration to be represented. In the combustion chamber, complete oxidation of the combustible fraction was assumed, producing mainly \(\mathrm{CO_2}\), \(\mathrm{H_2O}\), excess \(\mathrm{O_2}\), and inert ash. The outlet stream from this block was used to obtain the Aspen reference combustion temperature and product composition. These Aspen results were then compared with the simplified Python heat-balance model using the same empirical fuel formula, moisture and ash fractions, and stoichiometric ratio.

\begin{figure}[htbp]
    \centering
    \includegraphics[width=\textwidth]{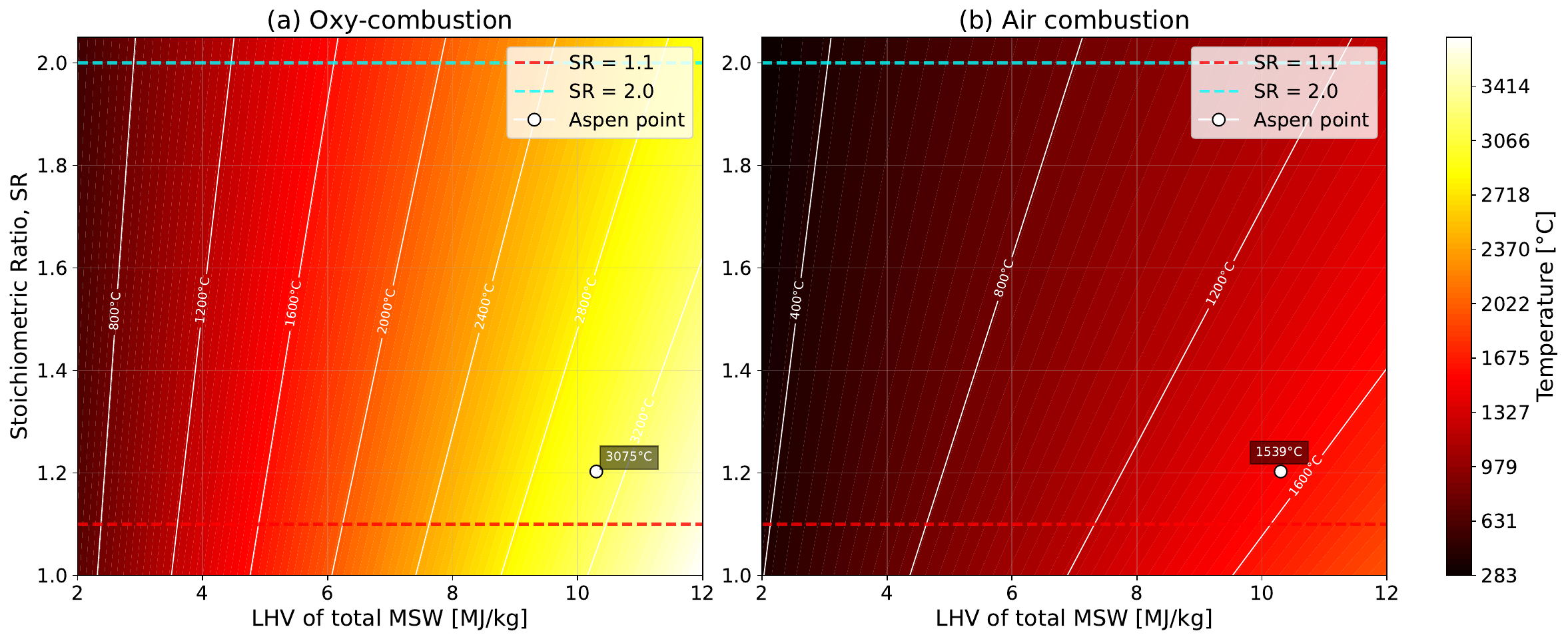}
    \caption{Calculated combustion-temperature maps for the MSW combustion process under (a) oxy-combustion and (b) air-combustion conditions. The maps show the combustion temperature as a function of the lower heating value of the total as-received MSW stream, \(LHV_{\mathrm{MSW}}\), and the stoichiometric ratio \(SR\). The comparison illustrates the strong increase in temperature achieved by excluding nitrogen from the oxidizing medium.}
    \label{fig:oxy_air_map}
\end{figure}

Figure~\ref{fig:oxy_air_map} shows the calculated combustion-temperature maps for the Aspen-derived MSW composition under oxy-combustion and air-combustion conditions. The maps are presented as functions of \(LHV_{\mathrm{MSW}}\) and \(SR\). A clear difference between the two combustion modes is observed. Under oxy-combustion conditions, the exclusion of nitrogen from the oxidizing medium substantially reduces the heat capacity of the product mixture, which results in significantly higher combustion temperatures. In contrast, in air combustion, nitrogen acts as a thermal ballast and lowers the attainable temperature over the full range of \(LHV_{\mathrm{MSW}}\) and \(SR\).

This result demonstrates that oxygen-based combustion can provide autothermal operation for a much wider range of waste compositions, including low-calorific-value and high-moisture wastes, without the need for additional support fuel. Removing nitrogen from the combustion process expands the operating window in which waste can sustain combustion by its own energy content. The air-combustion map shows that for low values of \(LHV_{\mathrm{MSW}}\), the combustion temperature may become too low for stable operation, whereas the oxy-combustion map remains in a much higher temperature range.

At the same time, the very high temperatures predicted for oxy-combustion without flue gas recirculation should be interpreted only as illustrative values that demonstrate the thermodynamic potential of the process. Such temperatures are not intended as practical operating targets. In real operation, the combustion temperature should be controlled to a substantially lower range, typically around \(1350\)--\(1400\,^{\circ}\mathrm{C}\), in order to limit excessive thermal loading of the reactor, corrosion, ash-related problems, and deterioration of construction materials. Therefore, the main role of the no-recirculation temperature map is to show why oxygen-based combustion makes autothermal treatment of low-grade waste possible, rather than to define the final design temperature of the process.

\begin{figure}[htbp]
    \centering
    \includegraphics[width=\textwidth]{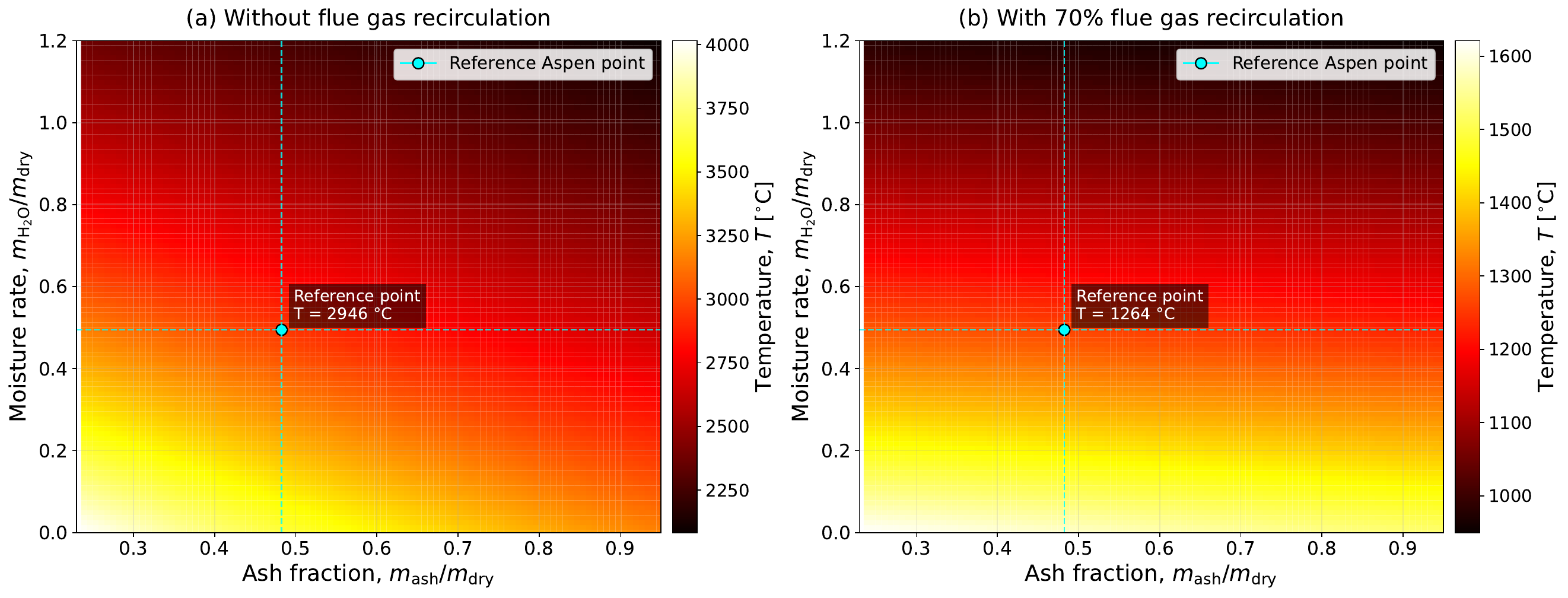}
    \caption{Aspen-based combustion-temperature maps showing the effect of ash fraction and moisture rate on oxy-combustion temperature: (a) without flue gas recirculation and (b) with \(70\%\) flue gas recirculation. The marked reference point corresponds to the Aspen reference waste composition. Increasing moisture and ash content lowers the combustion temperature, while flue gas recirculation shifts the process from the illustrative high-temperature regime toward a practical operating range.}
    \label{fig:ash_moisture_combined}
\end{figure}

The influence of waste composition and flue gas recirculation is shown more explicitly in Figure~\ref{fig:ash_moisture_combined}. The figure presents Aspen-based combustion-temperature maps as functions of ash fraction and moisture rate. Panel~(a) corresponds to oxy-combustion without flue gas recirculation, while panel~(b) corresponds to oxy-combustion with \(70\%\) flue gas recirculation. In both cases, increasing the moisture rate decreases the combustion temperature because part of the released heat is consumed by water heating and evaporation. Increasing ash content also lowers the temperature, since the ash behaves as an inert solid heat sink that must be heated together with the gaseous combustion products.

The comparison between the two panels also demonstrates the role of flue gas recirculation as a temperature-control mechanism. Without recirculation, the reference Aspen point corresponds to a very high combustion temperature of approximately \(2946\,^{\circ}\mathrm{C}\). With \(70\%\) flue gas recirculation, the reference temperature decreases to approximately \(1264\,^{\circ}\mathrm{C}\). Thus, the same waste composition can move from an illustrative high-temperature thermodynamic limit to a more practical operating range when recycled flue gas is introduced into the combustion chamber.

To further demonstrate the role of temperature control, the same calculation was repeated for oxy-combustion and air combustion with \(70\%\) flue gas recirculation. In this case, a major part of the flue gas stream was assumed to be cooled and recycled back to the combustion chamber. The purpose of this calculation was not to define the final engineering design of the reactor, but to show that the very high no-recirculation oxy-combustion temperatures can be reduced to a technically more realistic range by increasing the thermal ballast of the reacting mixture.

\begin{figure}[htbp]
    \centering
    \includegraphics[width=\textwidth]{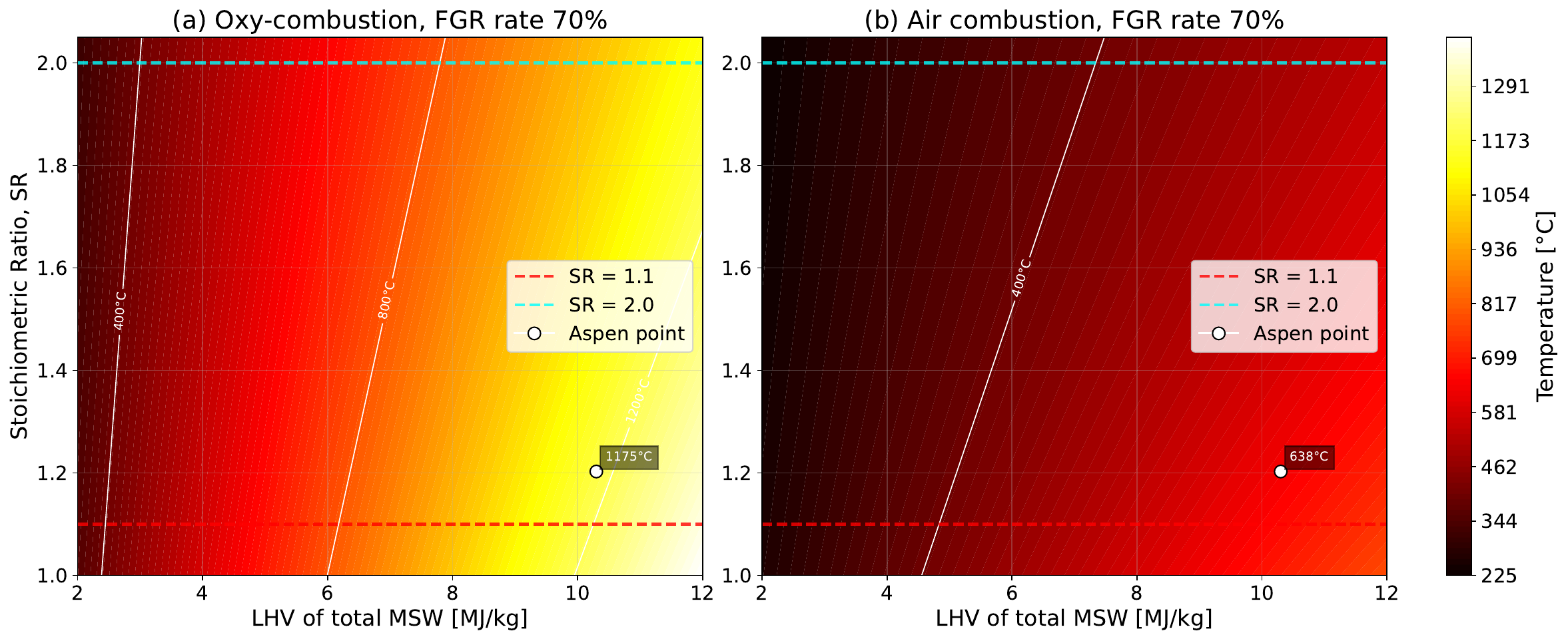}
    \caption{Calculated combustion-temperature maps for the Aspen-derived MSW composition under (a) oxy-combustion and (b) air-combustion conditions with \(70\%\) flue gas recirculation. The maps show the combustion temperature as a function of \(LHV_{\mathrm{MSW}}\) and stoichiometric ratio \(SR\). Compared with the no-recirculation case, FGR substantially reduces the oxy-combustion temperature while preserving the thermal advantage of oxygen-based combustion over air combustion.}
    \label{fig:oxy_air_map_fgr70}
\end{figure}

Figure~\ref{fig:oxy_air_map_fgr70} shows the combustion-temperature maps for oxy-combustion and air combustion at an FGR rate of \(70\%\). At the selected operating point, the oxy-combustion temperature decreases to approximately \(1175\,^{\circ}\mathrm{C}\) in the Python calculation, while the corresponding air-combustion temperature is approximately \(638\,^{\circ}\mathrm{C}\). Thus, even with strong flue gas recirculation, oxygen-based combustion remains thermally more favourable than air combustion because nitrogen dilution is avoided.

The simplified Python model and Aspen Plus simulations show the same qualitative behaviour over the considered range of flue gas recirculation rates. In both approaches, oxy-combustion without recirculation produces very high combustion temperatures, while increasing flue gas recirculation progressively reduces the temperature by increasing the heat capacity of the reacting mixture. The detailed numerical comparison for \(0\%\), \(30\%\), \(50\%\), and \(70\%\) FGR is provided in Appendix A, Supplementary Table~S2. The deviation between the Python and Aspen oxy-combustion temperatures remains below approximately \(8\%\) when calculated relative to the temperature rise above room temperature, which is acceptable for a generic feasibility-level model.

The no-recirculation case should therefore be interpreted as an upper thermodynamic limit rather than as a proposed operating regime. Its purpose is to demonstrate the fundamental effect of nitrogen removal: when the oxidizer does not contain atmospheric nitrogen, the heat released by waste oxidation is distributed over a smaller amount of product gas, and the attainable temperature rises substantially. This makes oxygen-based combustion suitable for autothermal treatment of low-calorific-value and high-moisture waste streams that may require additional support fuel under conventional air-combustion conditions.

At the same time, such high temperatures are not desirable for continuous industrial operation. Excessive combustion temperatures may intensify corrosion, accelerate degradation of refractory and metallic components, increase ash melting and slagging effects, and promote volatilization of metals from the inorganic fraction of waste. Therefore, the proposed process should be considered not as unrestricted high-temperature combustion, but as an oxygen-based thermal treatment system with deliberate temperature control. Flue gas recirculation is the principal control mechanism in this scheme.

At \(70\%\) flue gas recirculation, both calculation approaches predict that the oxy-combustion temperature can be shifted toward the \(1200\)--\(1300\,^{\circ}\mathrm{C}\) range for the selected reference waste composition. This range is much closer to the practically relevant operating window than the no-recirculation limit. Further adjustment of the recirculation rate, oxygen excess, oxidizer preheating, and waste moisture content can be used to approach the target range of approximately \(1350\)--\(1400\,^{\circ}\mathrm{C}\), where high-temperature oxidation and ash transformation can be achieved while avoiding unnecessarily severe thermal loading of the reactor.

Overall, the thermodynamic analysis leads to three main conclusions. First, exclusion of nitrogen from the combustion atmosphere substantially increases the attainable combustion temperature and expands the range of waste compositions that can be treated autothermally. Second, the simplified Python heat-balance model agrees reasonably well with Aspen Plus and can therefore be used for parametric screening of \(LHV_{\mathrm{MSW}}\), stoichiometric ratio, moisture content, ash content, and flue gas recirculation. Third, flue gas recirculation is essential for practical oxy-combustion of MSW, because it allows the process temperature to be reduced from the theoretical high-temperature limit to a controllable industrial operating range.

\section{Exhaust filtration with pure oxygen burning}\label{sec5}
When nitrogen is excluded from the waste incineration process, water vapor occupies a significantly larger proportion of the combustion gases than during combustion in atmospheric air. For the waste composition under consideration, the calculated partial pressure of water vapor at a total pressure of 1 bar is approximately 0.178 bar for combustion in air and 0.526 bar for oxy-fuel combustion, i.e., the increase in water partial pressure is approximately threefold.

The onset of condensation with decreasing temperature is determined by the ratio between the partial pressure of water vapor \(p_{\mathrm{H_2O}}\) and the saturated vapor pressure of water \(p_{\mathrm{sat}}(T)\). As the flue gas cools, condensation becomes thermodynamically possible when

$$ p_{\mathrm{H_2O}}>p_{\mathrm{sat}}(T). $$

The temperature at which

$$ p_{\mathrm{H_2O}}=p_{\mathrm{sat}}(T_{\mathrm{dew}})$$ 
corresponds to the dew point. Under the conditions under consideration, it is approximately 57.5°C for combustion in air and 82.6°C for combustion in oxygen. It can be seen that in the oxyfuel case, condensation begins at a significantly higher temperature, which expands the temperature range within which latent heat recovery and condensation cleaning of the flue gas can simultaneously occur, reducing energy costs.
For quantitative assessment, the condensation efficiency $\eta_c$ is defined as
the fraction of inlet water vapour transferred to the liquid phase during cooling.
For an approximately isobaric process at total pressure $P$, the molar flow rate
of the dry-gas components remains unchanged during condensation. Therefore, the
water-to-dry-gas molar ratio can be expressed directly in terms of the water-vapour
partial pressure as $p_{\mathrm{H_2O}}/(P-p_{\mathrm{H_2O}})$. The condensation
efficiency can thus be written as

\begin{equation}
\eta_c
=
\frac{
P\left[p_{\mathrm{H_2O,in}}-p_{\mathrm{sat}}(T_{\mathrm{cool}})\right]
}{
p_{\mathrm{H_2O,in}}
\left[P-p_{\mathrm{sat}}(T_{\mathrm{cool}})\right]
}.
\label{eq:cond_eff_pressure}
\end{equation}

When \(p_{\mathrm{H_2O}}\le p_{\mathrm{sat}}(T_{\mathrm{cool}})\)\, \(\eta_c=0\) is assumed. The resulting dependence is shown in Fig.~5(a).
\begin{figure}[htb]
 \centering
 \includegraphics[width=0.95\linewidth]{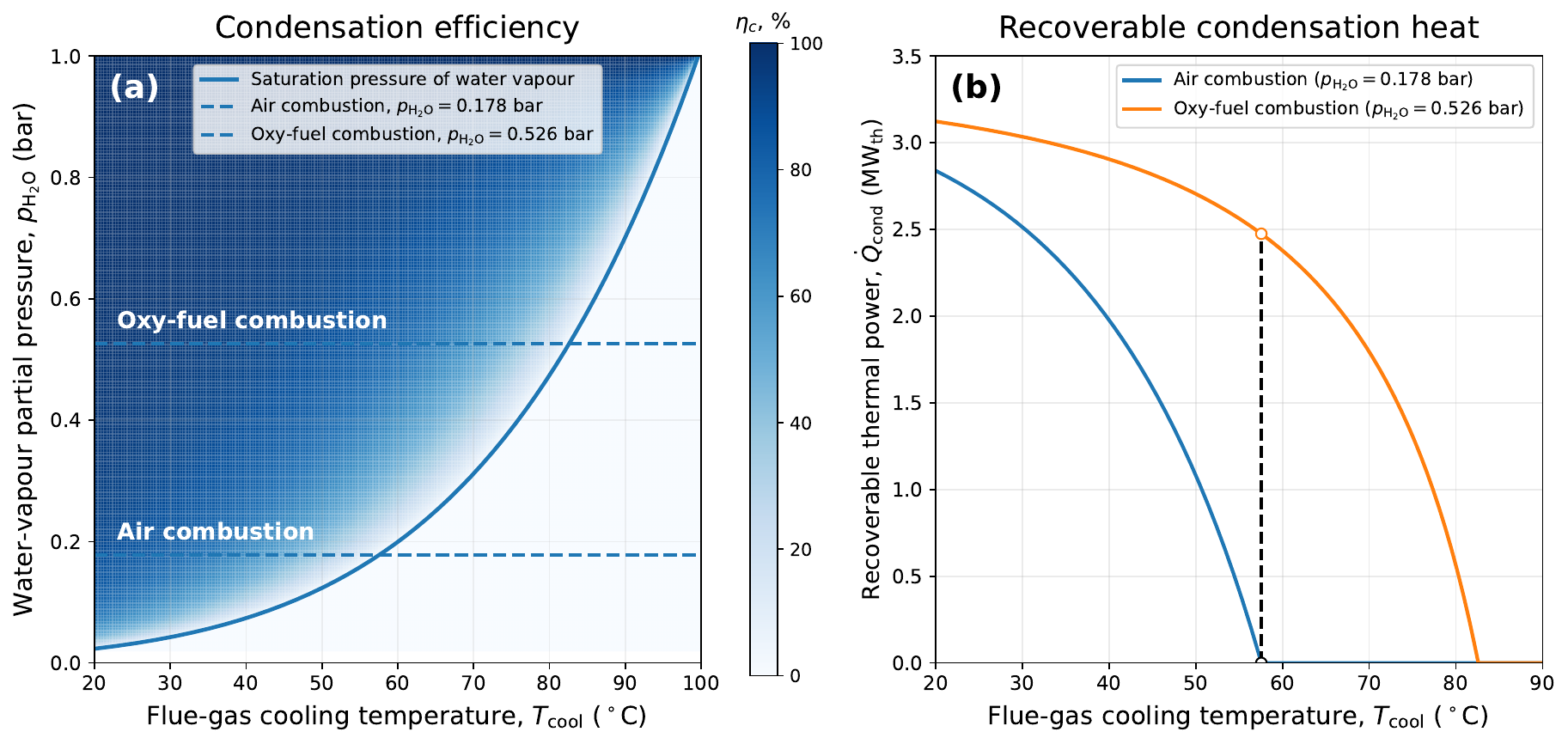}
\caption{Condensation behaviour of the flue gas for air and oxy-fuel combustion.
(a) Condensation efficiency as a function of cooling temperature and water-vapour
partial pressure. The saturation pressure of water was calculated using CoolProp,
and the dashed horizontal lines indicate the calculated water-vapour partial pressures
for air combustion ($p_{\mathrm{H_2O}}=0.178$~bar) and oxy-fuel combustion
($p_{\mathrm{H_2O}}=0.526$~bar).
(b) Recoverable thermal power from water condensation for the same two cases.
The vertical dashed line marks the air-combustion dew point
($T_{\mathrm{dew}}\approx57.5\,^{\circ}\mathrm{C}$) and extends to the corresponding
oxy-fuel curve, illustrating that substantial water condensation and latent-heat
recovery remain possible in the oxy-fuel case at a temperature where condensation
has already ceased in the air-combustion case.}
 \label{fig:5}
\end{figure}
The amount of heat that can be recovered during water vapor condensation is estimated as:

\begin{equation}
\dot Q_{\mathrm{cond}}
=
\dot m_{\mathrm{H_2O,in}}\,
\eta_c\,
h_{\mathrm{fg}}(T_{\mathrm{cool}}),
\label{eq:condensation_heat}
\end{equation}

where \(h_{\mathrm{fg}}\) is the specific heat of vaporization of water.
As shown in Fig.~5(b), oxy-fuel flue gas provides significantly greater heat
recovery potential. At the air-combustion dew point of approximately
\(57.5\,^{\circ}\mathrm{C}\), condensation in the air-fired case is only
beginning, whereas approximately \(2.5\,\mathrm{MW_{th}}\) of condensation heat
can be recovered from the oxy-fuel flue gas.

In the proposed process, this recovered low-temperature heat is used for process
preheating. For the present screening-level energy balance, it is assumed that
\(30\%\) of the recovered condensation heat contributes to an increase in the
electric output of the integrated plant. This corresponds to an additional
electric power of approximately
\[
P_{\mathrm{cond,int}}
=
0.30 \times 2.5
=
0.75\,\mathrm{MW_e}.
\]

Suspended ash particles can serve as heterogeneous condensation centers. To evaluate the process, we will use the lower limit of the experimental range of center concentrations of \(N=10^{13}\,\mathrm{m^{-3}}\) \cite{bergman2026physicochemical}. At \(T_{\mathrm{cool}}=70^\circ\mathrm{C}\) and \(p_{\mathrm{H_2O}}=0.526\) bar, approximately 59\% of the water vapor condenses. If the resulting condensate is distributed between these particles, the characteristic droplet diameter is approximately \(4~\mu\mathrm{m}\), and the specific gas-liquid interface area

\begin{equation}
a=4\pi Nr^2.
\label{eq:interfacial_area}
\end{equation}
reaches the order of

$$ a\sim500~\mathrm{m^{-1}}. $$

At the initial stage, droplet growth can be approximately considered as diffusion-limited transfer of water vapor from the gas volume to the surface of a spherical droplet:

\begin{equation}
\frac{dr}{dt}=\frac{K}{r}.
\label{eq:droplet_growth}
\end{equation}

where \(r\) is the droplet radius, and the growth coefficient \(K\) has the dimensions \(\mathrm{m^2\,s^{-1}}\) and is defined as

\begin{equation}
K=
\frac{D_{\mathrm{H_2O}}c_gM_w}{\rho_l}
\ln\!\left(
\frac{1-y_{w,s}}
{1-y_{w,\infty}}
\right).
\label{eq:growth_coefficient}
\end{equation}

Here, \(D_{\mathrm{H_2O}}\) is the diffusion coefficient of water vapor in a CO\(_2\)-rich gas mixture, \(c_g\) is the total molar concentration of the gas, \(M_w\) is the molar mass of water, \(\rho_l\) is the density of liquid water, \(y_{w,\infty}=p_{\mathrm{H_2O}}/P\) is the mole fraction of water vapor in the main gas flow, and \(y_{w,s}=p_{\mathrm{sat}}(T_d)/P\) is its equilibrium mole fraction at the droplet surface with temperature \(T_d\). The logarithmic term characterizes the driving force of condensation: the higher the water content in the main flow relative to the saturated state at the droplet surface, the faster the droplet grows.

Integrating this expression at approximately constant \(K\) yields

\begin{equation}
r^2-r_0^2=2Kt.
\label{eq:droplet_growth_integrated}
\end{equation}

where \(r_0\) is the initial radius of the particle or droplet, and \(t\) is the time. For the oxyfuel case under consideration, a purely diffusion estimate yields a growth time to micron size of the order of \(10^{-4}-10^{-3}\) s.

However, this estimate assumes that the droplet temperature remains close to the gas temperature. In reality, condensation is accompanied by the release of latent heat, causing the droplet to heat up. Its temperature balance can be written as

\begin{equation}
m_d c_{p,l}\frac{dT_d}{dt}
=
L_v\frac{dm_d}{dt}
-
hA_d(T_d-T_g).
\label{eq:droplet_energy_balance}
\end{equation}

where \(m_d\) and \(A_d\) are the mass and surface area of the droplet, \(c_{p,l}\) is the heat capacity of liquid water, \(L_v\) is the latent heat of condensation, \(h\) is the heat transfer coefficient between the droplet and the gas, and \(T_g\) is the temperature of the main gas flow. Increasing \(T_d\) increases \(p_{\mathrm{sat}}(T_d)\), reduces the difference between the water concentration in the gas and the equilibrium concentration at the surface, and thereby slows down further condensation. For the chosen calculation case, a combined assessment of mass transfer and heat removal yields a characteristic droplet growth time of the order of

$$ {\tau_{\mathrm{growth}}\sim10^{-2}\ \mathrm{s}}.$$

After droplet formation, the large interfacial surface area ensures rapid transfer of soluble components from the gas. For component \(i\), the gas-phase mass transfer rate can be estimated as

\begin{equation}
R_i=k_{g,i}a(c_i-c_i^*),
\qquad
k_{g,i}\simeq\frac{D_i}{r}.
\label{eq:gas_mass_transfer}
\end{equation}

where \(D_i\) is the diffusion coefficient of the component in the gas mixture. For \(r\sim2~\mu\mathrm{m}\), \(a\sim500~\mathrm{m^{-1}}\), and \(D_i\sim10^{-5}~\mathrm{m^2/s}\), the characteristic time for the delivery of molecules to the droplet surface,

\begin{equation}
\tau_i\sim\left(k_{g,i}a\right)^{-1}.
\label{eq:mass_transfer_time}
\end{equation}

is on the order of \(10^{-4}-10^{-3}\) s. Thus, the gas-phase diffusion of HCl, SO\(_2\), and other soluble impurities to the droplet surface occurs significantly faster than the growth of the droplets themselves. However, this time characterizes only the transport to the surface; the overall removal efficiency is also determined by dissolution, acid-base reactions, ash composition, and the residence time of the gas in the apparatus.
Representative gas-phase diffusion coefficients at \(70\,^{\circ}\mathrm{C}\),
together with literature data on interfacial uptake, aqueous behaviour, and
the expected condensate capture of selected flue-gas impurities, are summarized
in Appendix A, Supplementary Table S4.
The formation of an acidic condensate also creates a material-selection
constraint. The most aggressive conditions are expected near the beginning
of condensation, where the liquid phase can contain dissolved chlorides,
sulfur species, fluorides, and other impurities at elevated temperature.
A preliminary screening of candidate construction materials for the
condensation section is provided in Appendix A, Supplementary Table S5.
The separated condensate contains captured ash particles and dissolved gas impurities is sent to sedimentation tanker. Sediment composed of fly ash particles is sent combustion chamber for melting. Remaining water can be additionally purified and used for technical or agricultural needs.

\FloatBarrier

\section{CCUS potential of the oxy-fuel exhaust stream}\label{sec6}
A key advantage of oxy-fuel MSW combustion is the formation of a concentrated carbon-containing exhaust stream. In conventional air-fired waste incineration, the flue gas is strongly diluted by atmospheric nitrogen, which increases the total gas flow rate and reduces the partial pressure of \(\mathrm{CO_2}\). In the proposed oxygen-based process, nitrogen dilution is largely avoided. The main gaseous products of complete combustion are therefore \(\mathrm{CO_2}\), \(\mathrm{H_2O}\), residual \(\mathrm{O_2}\), and minor impurity species associated with the waste composition \citep{11283917}.

After heat recovery and cooling, a substantial fraction of water vapour can be removed by condensation. This step simultaneously reduces the water content of the exhaust stream and increases the relative concentration of \(\mathrm{CO_2}\) in the remaining gas phase. Therefore, the outlet gas after cooling and condensation can be considered a \(\mathrm{CO_2}\)-rich stream suitable for further purification, compression, storage, or utilization. This is one of the central differences between the proposed oxy-fuel configuration and conventional air-fired WtE plants with post-combustion capture.

The \(\mathrm{CO_2}\)-rich gas should not, however, be interpreted as a final storage-grade product without additional conditioning. Depending on the intended CCUS route, further treatment may be required to remove residual water, excess oxygen, acid gases, trace organic compounds, and other impurities. Additional compression or liquefaction may also be necessary for transport, geological storage, or industrial utilization. Thus, the oxy-fuel process simplifies the formation of a concentrated \(\mathrm{CO_2}\)-containing stream, but it does not eliminate the need for downstream gas conditioning.

The climate relevance of CCUS in waste-to-energy systems is linked to the mixed origin of carbon in MSW. Part of the carbon is fossil-derived, for example from plastics and synthetic materials, while another part is biogenic, originating from paper, food waste, wood, textiles, and other biomass-derived fractions. Capturing \(\mathrm{CO_2}\) from the fossil fraction reduces direct fossil greenhouse-gas emissions. Capturing and permanently storing \(\mathrm{CO_2}\) from the biogenic fraction can potentially create negative emissions, depending on the carbon-accounting framework and storage permanence.

In this generic feasibility study, the detailed design of the \(\mathrm{CO_2}\) purification and compression train is outside the main scope. Instead, the downstream CCUS stage is represented by a specific electricity demand per tonne of captured \(\mathrm{CO_2}\), which is included in the energy-balance calculation in Section 8. This approach allows the main effect of CCUS on the net power balance to be estimated while keeping the focus on the integrated oxy-fuel combustion, heat-recovery, and ASU system.

\section{Conceptual scheme of the plant}\label{sec7}

The burning temperature above $1300,^{\circ}\mathrm{C}$ makes the utilisation of a standard combustion chamber with a travelling grate impossible. Instead, a molten slag combustion chamber \cite{fang2026pilot} can be used. In this case, instead of toxic ash, rock-like materials are produced by the incineration plant.

Note that condensation filtering provides purification of FG from dust, HCl, HF, SO$_2$, SO$_3$, and N$_2$O without consumable materials like fabric filters, activated carbon, etc. Fly ash captured by water will experience sedimentation, and the sediment can be returned to the slag bath combustion chamber for melting.

\begin{figure}[htbp]
    \centering
    \includegraphics[width=\textwidth]{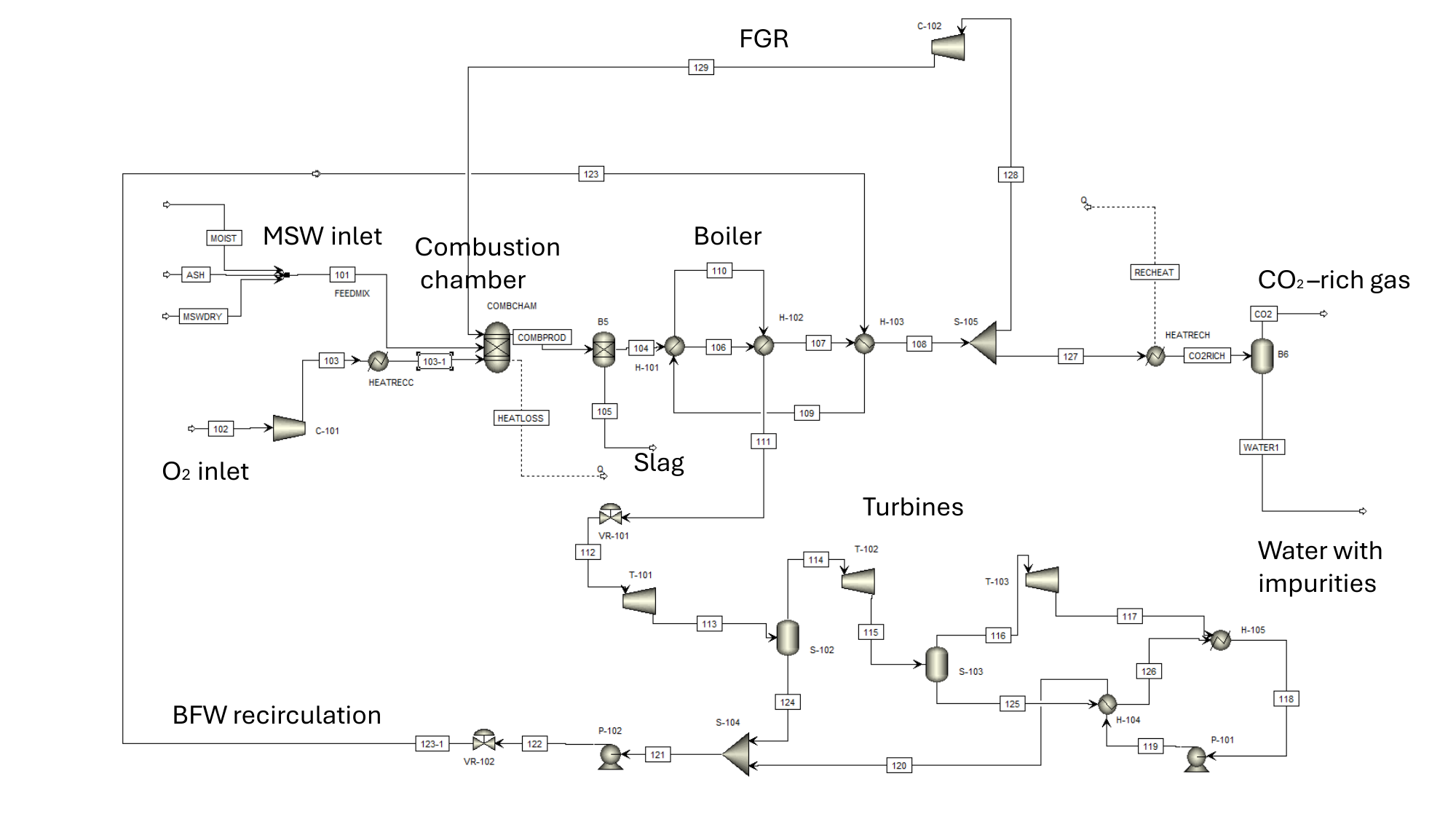}
    \caption{Conceptual Aspen Plus flowsheet of the proposed oxy-fuel municipal solid waste incineration plant. 
    The scheme includes MSW and oxygen inlet streams, the combustion chamber, slag withdrawal, boiler heat recovery, multi-stage steam expansion for power generation, boiler feed water recirculation, flue gas recirculation for temperature control, and downstream cooling and separation of the exhaust stream into condensed water with impurities and \ch{CO2}-rich gas.}
    \label{fig:aspen_flowsheet}
\end{figure}

The conceptual plant scheme used in the Aspen Plus simulation is shown in Fig.~\ref{fig:aspen_flowsheet}. 
It represents an oxy-fuel municipal solid waste incineration plant with flue gas recirculation (FGR), heat recovery in a steam cycle, and downstream cooling and separation of the exhaust stream. 
The scheme follows a conventional waste-to-energy layout with steam generation, but the air-fed grate furnace is replaced by an oxygen-fed combustion chamber. 
This reduces nitrogen dilution and produces a flue gas stream enriched in carbon dioxide and water vapour.

The MSW feed is represented by three inlet streams: dry combustible matter, moisture, and inert ash. 
These streams are mixed before entering the combustion chamber, while oxygen is supplied through a separate inlet. 
Inside the chamber, the combustible fraction is assumed to undergo complete oxidation, producing mainly \ch{CO2} and \ch{H2O}, with residual \ch{O2} depending on the stoichiometric ratio. 
The inorganic fraction is treated as inert ash and leaves the high-temperature zone as slag.

The hot combustion products are directed to the boiler section, where sensible heat is recovered for steam generation. 
The generated steam is expanded in a multi-stage turbine system for power production, then condensed and returned through the boiler feed water recirculation line. 
This part of the scheme represents the energy-recovery function of the plant and follows the general structure of conventional waste-to-energy steam cycles.

A fraction of the flue gas stream is separated after heat recovery and recycled to the combustion chamber. 
This FGR loop provides thermal moderation: recycled flue gas acts as a thermal ballast and shifts the oxy-fuel combustion temperature toward a technically acceptable operating range.

The remaining flue gas is sent to downstream cooling and separation. 
Because nitrogen is largely excluded, the exhaust contains high partial pressures of \ch{CO2} and water vapour. 
Cooling causes water condensation; the condensed phase is expected to remove dust and soluble acidic impurities, while the gas phase becomes enriched in \ch{CO2}. 
Thus, the final outlet streams are represented as water with impurities and \ch{CO2}-rich gas suitable for further purification, compression, storage, or utilization.

Overall, the proposed scheme integrates oxygen-based combustion of heterogeneous waste, high-temperature transformation of mineral residue into slag, heat recovery through a steam cycle, and production of a concentrated \ch{}{CO2}-containing exhaust stream. 
This supports improved heat recovery, condensation-based gas cleaning, and simplified downstream carbon capture, while FGR provides temperature control for practical operation.

\FloatBarrier

\section{Energy balance, ASU penalty, and power available for CCUS}\label{sec8}

The energetic feasibility of the proposed oxy-fuel MSW process was evaluated using the Aspen Plus flowsheet shown in Figure~\ref{fig:aspen_flowsheet}. The electricity-producing part of the process is represented by the steam-turbine cycle, while the main electricity-consuming units include the oxygen-line compressor C-101, the flue-gas-recirculation compressor C-102, and the pumps P-101 and P-102 in the water/steam cycle. In addition to these internal loads, the analysis includes the electricity demand of oxygen production in the air separation unit (ASU) and an estimated electricity demand for downstream \(\mathrm{CO_2}\) conditioning and CCUS.

The energy balance was evaluated for the representative flue-gas-recirculation case with a combustion temperature of \(1350\,^{\circ}\mathrm{C}\). Although combustion temperature is not explicit in the net-power equation, it affects the oxygen excess required for complete oxidation. This temperature was selected because it lies within the practically relevant high-temperature operating range discussed in the thermodynamic analysis. At this operating point, flue gas recirculation provides temperature control while preserving the thermal advantage of oxy-fuel combustion.

The gross electric power generated by the steam cycle was calculated as
\begin{equation}
P_{\mathrm{gross}}
=
W_{\mathrm{turb1}}+
W_{\mathrm{turb2}}+
W_{\mathrm{turb3}},
\end{equation}
where \(W_{\mathrm{turb1}}\), \(W_{\mathrm{turb2}}\), and \(W_{\mathrm{turb3}}\) are the power outputs of the three turbine stages. For the selected \(1350\,^{\circ}\mathrm{C}\) FGR case, the gross turbine output was
\begin{equation}
P_{\mathrm{gross}}
=
3734\ \mathrm{kW}.
\end{equation}

The internal auxiliary power demand was estimated as
\begin{equation}
P_{\mathrm{aux}}
=
W_{\mathrm{C101}}+
W_{\mathrm{C102}}+
W_{\mathrm{P101}}+
W_{\mathrm{P102}},
\end{equation}
where \(W_{\mathrm{C101}}\) is the oxygen-compressor demand, \(W_{\mathrm{C102}}\) is the flue-gas-recirculation compressor demand, and \(W_{\mathrm{P101}}\) and \(W_{\mathrm{P102}}\) are pump loads. For the selected operating point, the total auxiliary demand was
\begin{equation}
P_{\mathrm{aux}}
=
213\ \mathrm{kW}.
\end{equation}

The electricity demand of oxygen production was calculated from the fresh oxygen mass flow rate:
\begin{equation}
P_{\mathrm{ASU}}
=
e_{\mathrm{O_2}}\dot m_{\mathrm{O_2}},
\end{equation}
where \(\dot m_{\mathrm{O_2}}\) is the oxygen flow rate and \(e_{\mathrm{O_2}}\) is the specific electricity consumption of oxygen production. In this assessment,
\begin{equation}
e_{\mathrm{O_2}}
=
0.265~\mathrm{kWh\,kg^{-1}_{O_2}}
\end{equation}
was used. This value corresponds to \(265~\mathrm{kWh\,t^{-1}_{O_2}}\) and was selected as a commercially grounded VSA oxygen-production estimate. It should be interpreted as the ASU electricity penalty per unit mass of produced oxygen, rather than as a universal constant, because the specific energy demand of oxygen production depends on plant scale, oxygen purity, delivery pressure, and ASU technology.

For the selected case, the fresh oxygen flow rate was
\begin{equation}
\dot m_{\mathrm{O_2}}
=
8231\ \mathrm{kg\,h^{-1}}.
\end{equation}
The corresponding ASU electricity demand was therefore
\begin{equation}
P_{\mathrm{ASU}}
=
0.265 \times 8231
\approx
2181\ \mathrm{kW}.
\end{equation}

The electric power remaining after internal loads and oxygen-production penalty was defined as
\begin{equation}
P_{\mathrm{avail}}
=
P_{\mathrm{gross}}
-
P_{\mathrm{aux}}
-
P_{\mathrm{ASU}}.
\end{equation}
For the selected operating point, this gives
\begin{equation}
P_{\mathrm{avail}}
=
3734
-
213
-
2181
=
1340\ \mathrm{kW}.
\end{equation}

For the CCUS estimate, the \(\mathrm{CO_2}\) generation rate was calculated from the Aspen-derived empirical formula of the dry combustible fraction,
\(\mathrm{C_1H_{1.51334}O_{0.51837}}\). Complete oxidation of one mole of this empirical fuel produces one mole of \(\mathrm{CO_2}\). Therefore, the \(\mathrm{CO_2}\) mass flow can be estimated from the dry combustible mass flow as
\begin{equation}
\dot m_{\mathrm{CO_2}}
=
\dot m_{\mathrm{dry}}
\frac{M_{\mathrm{CO_2}}}{M_{\mathrm{fuel}}},
\end{equation}
where \(\dot m_{\mathrm{dry}}\) is the mass flow rate of the dry combustible fraction, \(M_{\mathrm{CO_2}}\) is the molar mass of carbon dioxide, and \(M_{\mathrm{fuel}}\) is the molar mass of the empirical dry combustible formula. For the reference case,
\begin{equation}
\dot m_{\mathrm{dry}}
=
4186\ \mathrm{kg\,h^{-1}},
\qquad
M_{\mathrm{fuel}}
\approx
21.83\ \mathrm{g\,mol^{-1}},
\qquad
M_{\mathrm{CO_2}}
=
44.01\ \mathrm{g\,mol^{-1}}.
\end{equation}
Thus,
\begin{equation}
\dot m_{\mathrm{CO_2}}
=
4186
\frac{44.01}{21.83}
\approx
8.44\ \mathrm{t\,h^{-1}}.
\end{equation}

Assuming a specific electricity demand \citep{IPCC2005CCS} of
\begin{equation}
e_{\mathrm{CCUS}}
=
120~\mathrm{kWh\,t^{-1}_{CO_2}},
\end{equation}
the corresponding CCUS power demand is approximately
\begin{equation}
P_{\mathrm{CCUS}}
=
e_{\mathrm{CCUS}}\dot m_{\mathrm{CO_2}}
\approx
120 \times 8.44
\approx
1014\ \mathrm{kW}.
\end{equation}

The remaining net electric power after oxygen production and CCUS was therefore calculated as
\begin{equation}
P_{\mathrm{net,CCUS}}
=
P_{\mathrm{avail}}
-
P_{\mathrm{CCUS}}
=
1340
-
1014
=
326\ \mathrm{kW}.
\end{equation}
An additional contribution to the electric power balance can be obtained through
integration of the low-temperature condensation heat into the process. As discussed
in Section~5, approximately \(2.5\,\mathrm{MW_{th}}\) of condensation heat is
available under the selected oxy-fuel conditions. This heat is assumed to be used
for process preheating, reducing the amount of higher-grade heat required elsewhere
in the plant. For the present screening-level assessment, \(30\%\) of the recovered
condensation heat is assumed to appear as an equivalent increase in electric power
output. Thus,

\begin{equation}
P_{\mathrm{cond,int}}
=
0.30\,\dot{Q}_{\mathrm{cond}}
=
0.30\times 2.5
=
0.75\ \mathrm{MW_e}.
\end{equation}

The resulting net electric power is therefore

\begin{equation}
P_{\mathrm{net,int}}
=
P_{\mathrm{net,CCUS}}
+
P_{\mathrm{cond,int}}
=
326+750
=
1076\ \mathrm{kW}.
\end{equation}

\begin{figure}[htbp]
    \centering
    \includegraphics[width=0.78\textwidth]{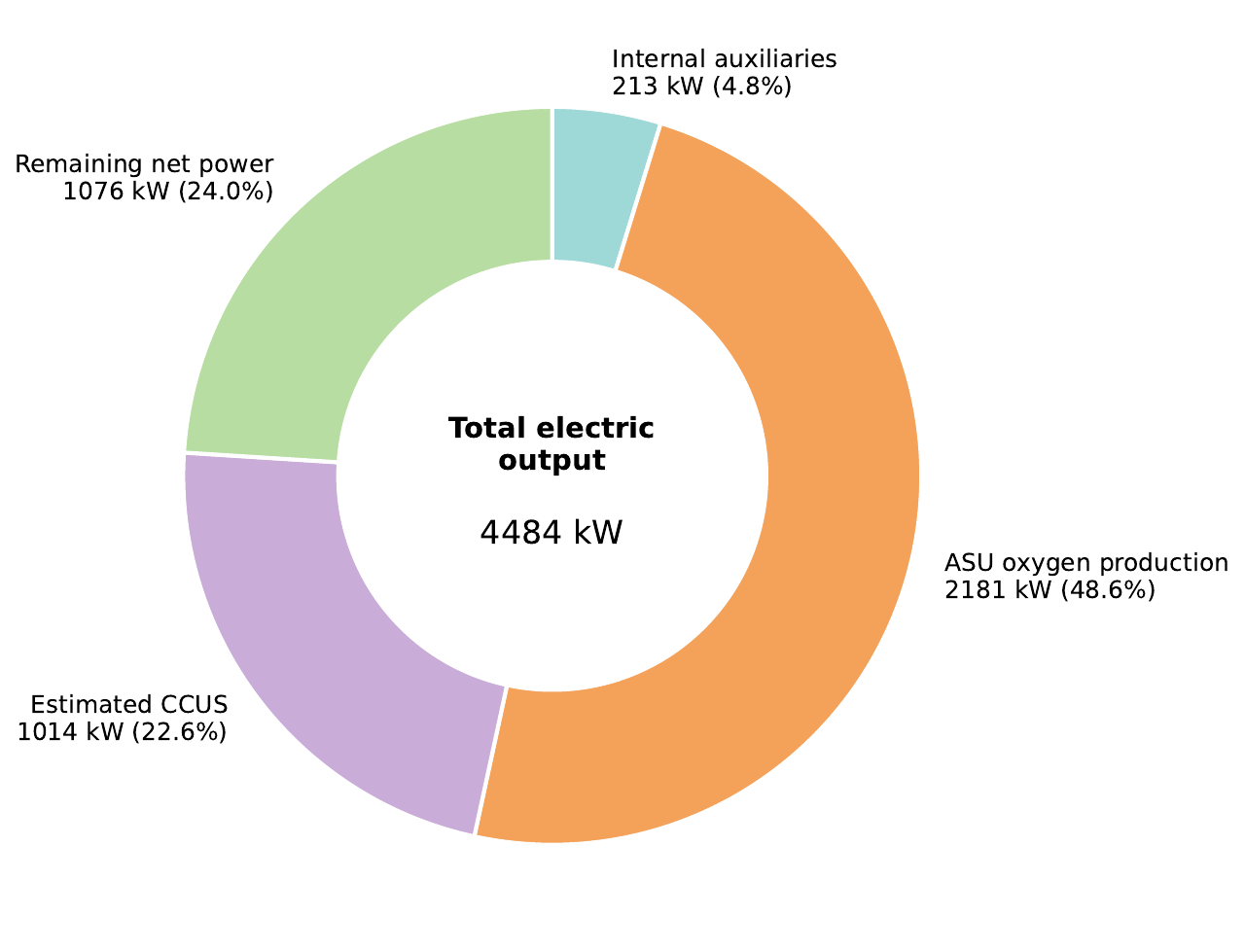}
    \caption{Distribution of total electric power for the oxy-fuel MSW process at a
target combustion temperature of \(1350\,^{\circ}\mathrm{C}\) with flue gas
recirculation. The total output includes the gross steam-cycle electric power
and an additional \(0.75\,\mathrm{MW_e}\) contribution attributed to process
preheating using recovered condensation heat, assuming that \(30\%\) of the
approximately \(2.5\,\mathrm{MW_{th}}\) available condensation heat contributes
to the electric power balance. The ASU electricity demand was calculated using
\(e_{\mathrm{O_2}}=0.265~\mathrm{kWh\,kg^{-1}_{O_2}}\). The resulting total
electric output is divided into internal auxiliary consumption, oxygen production
in the ASU, estimated CCUS electricity demand, and remaining net electric power.}
    \label{fig:energy_pie_1350}
\end{figure}

Figure~\ref{fig:energy_pie_1350} summarizes the distribution of the total electric
power for the selected \(1350\,^{\circ}\mathrm{C}\) FGR case, including the
estimated contribution from integration of the recovered condensation heat.
The largest share of the total electric output is consumed by oxygen production
in the ASU, which accounts for approximately \(48.6\%\). The estimated CCUS
demand accounts for approximately \(22.6\%\), while internal auxiliary loads
represent about \(4.8\%\). After subtracting these contributions, approximately
\(24.0\%\) of the total electric output remains as net power.

The result shows that the integrated FGR configuration retains a positive energy
margin even after accounting for oxygen production and an estimated CCUS penalty.
With the commercially grounded ASU estimate of
\(e_{\mathrm{O_2}}=0.265~\mathrm{kWh\,kg^{-1}_{O_2}}\), the remaining net
electric power before accounting for condensation-heat integration is approximately
\(326~\mathrm{kW}\). When the estimated \(0.75~\mathrm{MW_e}\) contribution
associated with the use of recovered condensation heat for process preheating is
included, the remaining net electric power increases to approximately
\(1.1~\mathrm{MW}\), corresponding to about \(24\%\) of the total electric
output. Thus, integration of the recovered condensation heat can substantially
improve the overall energy balance of the oxy-fuel MSW process.

The energy distribution highlights the dominant role of the ASU penalty in the oxy-fuel configuration. Oxygen production is the largest electricity-consuming term in the balance, and therefore the energetic feasibility of the process depends strongly on the specific electricity consumption of the ASU. This value is scale- and technology-dependent: smaller modular oxygen-production systems generally have higher specific energy demand, while larger and better integrated systems may achieve lower values. Consequently, ASU selection and integration should be treated as central design parameters rather than as external auxiliary details.

Overall, the energy-balance analysis supports three conclusions. First, the selected oxy-fuel MSW configuration with flue gas recirculation can retain a positive electric power margin after internal auxiliary loads, oxygen production, and estimated CCUS demand are included. Second, the ASU is the dominant parasitic load and should be a central consideration in further process optimization and plant-scale assessment. Third, the positive remaining net power suggests that the recuperated oxy-fuel configuration can potentially support both oxygen supply and downstream carbon management within the same integrated process, although the margin is sensitive to the selected ASU technology and its specific energy consumption.

\FloatBarrier

\section{Plant scale, ASU energy penalty, and capital cost}\label{sec9}

The results of the energy-balance analysis show that oxygen production is the dominant parasitic load in the proposed oxy-fuel MSW process. Therefore, the feasibility of the concept depends not only on the combustion temperature, heat recovery, and CCUS demand, but also on the scale and performance of the air separation unit (ASU). The specific electricity demand of oxygen production, \(e_{\mathrm{O_2}}\), is not a universal constant. It depends on the oxygen-production technology, oxygen purity, delivery pressure, plant capacity, and whether product compression is included.

For a required oxygen mass flow rate, the ASU electricity demand is

\begin{equation}
P_{\mathrm{ASU}}
=
e_{\mathrm{O_2}}\dot m_{\mathrm{O_2}},
\label{eq:asu_power_scale}
\end{equation}

where \(P_{\mathrm{ASU}}\) is the electric power consumed by oxygen production, \(e_{\mathrm{O_2}}\) is the specific electricity demand of oxygen production, and \(\dot m_{\mathrm{O_2}}\) is the oxygen mass flow rate. This linear dependence means that even moderate changes in \(e_{\mathrm{O_2}}\) have a direct effect on the net energy balance of the plant.

For the reference case considered in this study, the fresh oxygen demand is

\begin{equation}
\dot m_{\mathrm{O_2}}
=
8231~\mathrm{kg\,h^{-1}},
\end{equation}

which corresponds to an installed oxygen-production capacity of

\begin{equation}
\dot m_{\mathrm{O_2,day}}
=
8231 \times 24
\approx
1.98\times 10^5~\mathrm{kg\,day^{-1}}
=
197.5~\mathrm{t\,day^{-1}}.
\label{eq:o2_day_capacity}
\end{equation}

The corresponding as-received MSW feed rate in the Aspen reference case is

\begin{equation}
\dot m_{\mathrm{MSW}}
=
8276~\mathrm{kg\,h^{-1}},
\end{equation}

or

\begin{equation}
\dot m_{\mathrm{MSW,day}}
=
8276 \times 24
\approx
200~\mathrm{t\,day^{-1}}.
\label{eq:msw_day_capacity}
\end{equation}

Thus, for the selected waste composition and stoichiometric ratio, the oxygen requirement is approximately one kilogram of oxygen per kilogram of as-received waste:

\begin{equation}
\frac{\dot m_{\mathrm{O_2}}}{\dot m_{\mathrm{MSW}}}
=
\frac{8231}{8276}
\approx
0.995
\approx
1.
\label{eq:o2_msw_ratio_actual}
\end{equation}

This practical near-unity ratio is useful for interpreting ASU scale in terms of waste-treatment capacity. The reference case represents a medium-scale municipal waste-to-energy facility serving roughly \(2\times 10^5\) people.   This estimate should be interpreted as an approximate scale indicator, because not all generated MSW is normally sent to incineration: the actual relation between plant capacity and served population depends on local sorting, reuse, recycling rates, and the residual-waste fraction available for thermal treatment.

\begin{figure}[htbp]
    \centering
    \includegraphics[width=0.90\textwidth]{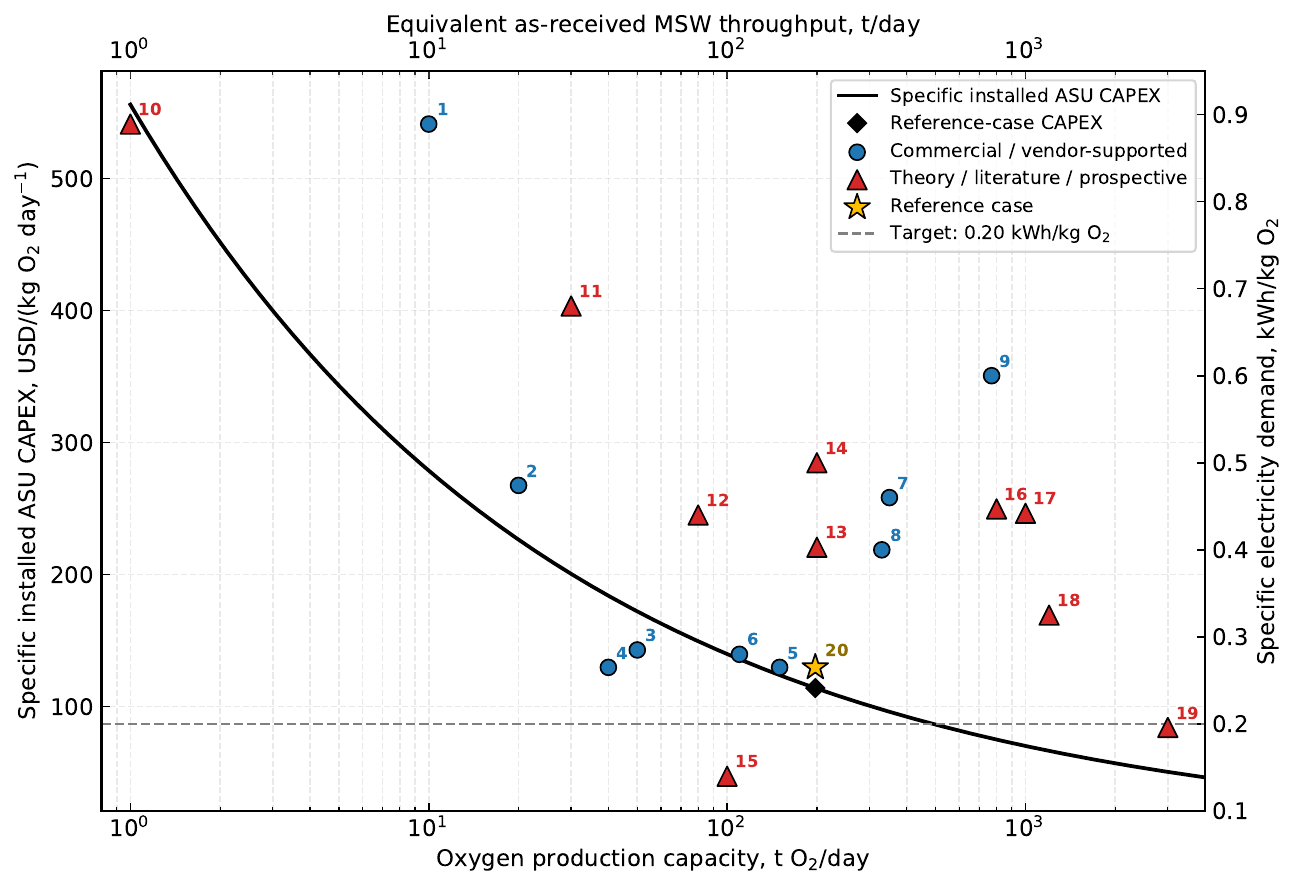}
    \caption{Specific electricity demand and capacity-normalized capital cost of oxygen production as functions of ASU scale. The lower horizontal axis shows oxygen production capacity, while the upper horizontal axis shows the approximately equivalent as-received MSW throughput for the reference waste composition used in this study. Commercial/vendor-supported ASU points are distinguished from literature, benchmark, or prospective oxygen-production cases.  The reference oxy-fuel MSW case corresponds to \(197.5~\mathrm{t\,day^{-1}}\) of oxygen production and approximately \(200~\mathrm{t\,day^{-1}}\) of as-received MSW throughput. The horizontal dashed line indicates the target value \(e_{\mathrm{O_2}}=0.20~\mathrm{kWh\,kg^{-1}_{O_2}}\), representing a favourable large-scale or optimized oxygen-production case. The numbering of data points corresponds to entries listed in Table S3 of the  Supplementary Materials, Appendix A, where detailed information and references for each case are provided.}
    \label{fig:asu_scale_capex}
\end{figure}

Figure~\ref{fig:asu_scale_capex} summarizes the relationship between ASU scale, specific electricity demand, and capital cost. The numerical points and literature or vendor sources used to construct the ASU-scale figure are listed in Appendix A, Supplementary Table S3. The lower horizontal axis shows the oxygen-production capacity, while the upper horizontal axis shows the approximately equivalent as-received MSW throughput. The right vertical axis shows the specific electricity demand of oxygen production, \(e_{\mathrm{O_2}}\), for commercial/vendor-supported ASU data and for literature, benchmark, or prospective oxygen-production cases. The left vertical axis shows the capacity-normalized ASU capital cost estimated from the scaling relation used in this study.

The base energy-balance calculation used

\begin{equation}
e_{\mathrm{O_2}}
=
0.265~\mathrm{kWh\,kg^{-1}_{O_2}},
\end{equation}

which gives

\begin{equation}
P_{\mathrm{ASU}}
=
0.265 \times 8231
\approx
2181~\mathrm{kW}.
\end{equation}

This value was selected as a commercially grounded ASU energy penalty. However, Figure~\ref{fig:asu_scale_capex} shows that \(e_{\mathrm{O_2}}\) can vary substantially depending on technology and scale. Smaller PSA or VPSA systems generally have higher specific electricity demand, while larger or more optimized systems can approach lower values. If the ASU specific electricity demand is reduced to the target value

\begin{equation}
e_{\mathrm{O_2}}
=
0.20~\mathrm{kWh\,kg^{-1}_{O_2}},
\end{equation}

then the ASU power demand for the same oxygen flow rate becomes

\begin{equation}
P_{\mathrm{ASU,target}}
=
0.20 \times 8231
\approx
1646~\mathrm{kW}.
\end{equation}

Compared with the base case, this would reduce the ASU electricity demand by approximately

\begin{equation}
2181 - 1646
=
535~\mathrm{kW}.
\end{equation}

Consequently, the remaining net electric power would increase from approximately
1.1 MW to approximately 1.6 MW when the ASU specific electricity demand is reduced
from 0.265 to 0.20 kWh kg$^{-1}_{\mathrm{O_2}}$.

The capital cost of the ASU was estimated using the scaling relation

\begin{equation}
C_{\mathrm{ASU}}
=
70\times 10^6
\left(
\frac{\dot m_{\mathrm{O_2,day}}}{10^6}
\right)^{0.7},
\label{eq:asu_capex}
\end{equation}

where \(C_{\mathrm{ASU}}\) is the estimated ASU capital cost in USD and \(\dot m_{\mathrm{O_2,day}}\) is the installed oxygen-production capacity in \(\mathrm{kg\,day^{-1}}\). The exponent below unity reflects economy of scale: larger plants require higher total capital investment, but the capital cost per unit of installed production capacity decreases \citep{Williams1947}.

To represent this effect, the capacity-normalized ASU capital cost was defined as

\begin{equation}
C_{\mathrm{ASU,spec}}
=
\frac{C_{\mathrm{ASU}}}{\dot m_{\mathrm{O_2,day}}}.
\label{eq:asu_specific_capex}
\end{equation}

The unit of \(C_{\mathrm{ASU,spec}}\) is

\begin{equation}
\mathrm{USD}/(\mathrm{kg_{O_2}\,day^{-1}}),
\end{equation}

which should be interpreted as the capital cost per installed unit of daily oxygen-production capacity, not as an operating cost per kilogram of oxygen produced.

For the reference case,

\begin{equation}
\dot m_{\mathrm{O_2,day}}
=
197500~\mathrm{kg\,day^{-1}},
\end{equation}

and therefore

\begin{equation}
C_{\mathrm{ASU}}
=
70\times 10^6
\left(
\frac{197500}{10^6}
\right)^{0.7}
\approx
22.5\times 10^6~\mathrm{USD}.
\end{equation}

The corresponding capacity-normalized capital cost is

\begin{equation}
C_{\mathrm{ASU,spec}}
=
\frac{22.5\times 10^6}{197500}
\approx
114~\mathrm{USD}/(\mathrm{kg_{O_2}\,day^{-1}}).
\end{equation}

This result shows that plant scale affects the ASU penalty in two ways. First, the specific electricity demand of oxygen production tends to be higher for small modular systems and lower for larger or better integrated systems. Second, the capacity-normalized capital cost decreases with increasing installed oxygen-production capacity according to the adopted scaling law. Therefore, small oxy-fuel MSW plants would be disadvantaged by both higher parasitic electricity demand and higher installed ASU cost per unit of oxygen capacity, whereas larger plants can benefit from scale effects.

Overall, the ASU should be treated as a central design component of the oxy-fuel MSW plant rather than as an external auxiliary system. The reference case remains energetically positive when \(e_{\mathrm{O_2}}=0.265~\mathrm{kWh\,kg^{-1}_{O_2}}\) is used, but the remaining power margin is limited. Reaching values closer to \(0.20~\mathrm{kWh\,kg^{-1}_{O_2}}\) would substantially improve the net energy balance. Thus, future optimization should consider the ASU, combustion chamber, flue gas recirculation loop, heat-recovery system, and CCUS unit as an integrated plant-scale system.
A preliminary economic assessment of the proposed process, including estimates
of operating revenues, air-separation-unit costs, and alternative product-utilization
scenarios, is provided in Appendix B.
\FloatBarrier

\section{Conclusion}\label{sec10}
This work presents a generic feasibility assessment of an integrated oxy-fuel MSW incineration concept combining oxygen-based combustion, flue gas recirculation, heat recovery, condensation-assisted exhaust treatment, and downstream CCUS. The results show that removing nitrogen from the oxidizing atmosphere substantially increases the attainable combustion temperature and expands the range of waste compositions that can be treated autothermally. This is especially relevant for low-calorific-value and high-moisture waste streams, which may require support fuel under conventional air combustion.

High combustion temperature ensures transformation of fly ash and slag into inert rock material.
The thermodynamic analysis also shows that uncontrolled oxy-combustion can lead to excessively high theoretical temperatures. Therefore, flue gas recirculation is a necessary temperature-control mechanism rather than only an optional improvement. In the reference Aspen case, recirculation shifts the process from an illustrative high-temperature limit toward a practical operating range suitable for continuous design.

The proposed process also changes exhaust-gas treatment. Because nitrogen dilution is largely avoided, the exhaust stream contains high partial pressures of water vapour and carbon dioxide. Cooling enables substantial water condensation, which can support condensation-assisted removal of dust and soluble impurities such as  \ch{SO2},  \ch{SO3},  \ch{HCl}, \ch{HF}, and\ch{NO2}. After water removal, the remaining gas phase is enriched in \ch{CO2}, making it more suitable for downstream purification, compression, storage, or utilization than nitrogen-diluted flue gas from conventional air-fired incineration.

The reference energy balance confirms that the integrated FGR configuration can retain a positive electric power margin after internal auxiliary loads, oxygen production, and an estimated CCUS penalty. The ASU is the dominant limiting subsystem: both its specific electricity demand and capacity-normalized capital cost depend strongly on plant scale and technology. The reference case corresponds to approximately 197.5 t/day of oxygen production and 200 t/day of as-received MSW throughput, representing a medium-scale municipal waste-treatment facility.

Overall, oxy-fuel MSW incineration with flue gas recirculation appears to be a technically coherent route toward high-temperature waste treatment, reduced nitrogen dilution, condensation-assisted exhaust cleaning, and simplified carbon management. At the same time, the present study remains a feasibility-level assessment. Future work should include detailed reactor modelling, ash and slag behaviour, corrosion and impurity chemistry, dynamic flue gas recirculation control, detailed \ch{CO2} purification design, and full techno-economic and life-cycle assessment.

\backmatter

\bmhead{Supplementary information}

Appendix A provides supplementary data on MSW properties, model comparison, ASU performance, condensation-assisted impurity removal, and preliminary material selection for the condensation section. Appendix B presents a preliminary economic assessment of the proposed process, including alternative air-separation and product-utilization scenarios.

\bmhead{Funding}

Funded by the European Union. Views and opinions expressed are however those of the authors only and do not necessarily reflect those of the European Union or the European Research Executive Agency (REA). Neither the European Union nor the granting authority can be held responsible for them.
This programme has received funding from the European Union through Marie Skłodowska-Curie actions under
project number 101081466: SEED - Systems and Engineering Science Doctorate.

\bmhead{Author Contribution}
Mikhail Kaliteevski: Conceptualization, project leadership, supervision,  development of the overall manuscript structure, and writing and critical revision  of the manuscript.  Leonid Chechurin: Literature review and analysis of the research and industrial  landscape, methodology, modelling support, and critical revision of the manuscript.  Elizaveta Girshova: Literature review and analysis of the research and industrial  landscape, Python modelling, data analysis, visualization, preparation of  figures, interpretation of results, and writing and revision of the manuscript.  Fedor Vasilyev: Aspen Plus modelling and simulations, analysis of simulation results,  methodology, and critical revision of the manuscript.  Maxim Permyakov: Scientific discussions, interpretation of results, conceptual input,  and critical review and editing of the manuscript.  All authors read and approved the final manuscript. 

\bmhead{Conflict of Interest}
On behalf of all authors, the corresponding author states that there is no conflict of interest.

\bmhead{Data Availability}

The data supporting the findings of this study are included in the article and its Supplementary Information. Additional calculation files and model data are available from the corresponding author upon reasonable request.







\clearpage
\appendix

\begin{center}
{\Large Appendix A. Supplementary tables}\par\vspace{0.3em}

\end{center}

\begin{landscape}
\setlength{\tabcolsep}{0.8pt}
\begin{ThreePartTable}
\begin{TableNotes}[flushleft]
\footnotesize
\item Notes: Values are approximate screening ranges, because ignition and sustained combustion depend on particle size, moisture, additives, geometry, heating rate, and ventilation. Ignition and fire-behaviour ranges were compiled from general fire-safety and combustion references \parencite{BabrauskasIgnitionHandbook,SFPEHandbook}; waste-composition ranges are representative municipal-solid-waste ranges informed by global waste-management data \parencite{Kaza2018WhatWaste}. The table is intended for qualitative comparison and should not replace material-specific safety data sheets or local waste-characterization measurements.
\end{TableNotes}
\scriptsize
\begin{longtable}{L{2.3cm} L{2.9cm} L{3.0cm} C{1.5cm} L{4.6cm} C{1.4cm} C{1.4cm} L{5.0cm}}
\caption{Major components of municipal solid waste, ignition behaviour in air, limiting oxygen index (LOI), expected behaviour in oxygen-enriched atmosphere, approximate content, lower heating value (LHV), and simplified composition.}\label{tab:S1_ignition}\tabularnewline
\toprule
Component & Ignition / autoignition in air, \si{\degreeCelsius} & Sustained combustion in air & LOI, vol.\% \ce{O2} & Behaviour in oxygen-enriched atmosphere & Content, \% & LHV, MJ kg$^{-1}$ & Chemical formula / composition \\
\midrule
\endfirsthead
\caption[]{Major components of municipal solid waste, continued.}\tabularnewline
\toprule
Component & Ignition / autoignition in air, \si{\degreeCelsius} & Sustained combustion in air & LOI, vol.\% \ce{O2} & Behaviour in oxygen-enriched atmosphere & Content, \% & LHV, MJ kg$^{-1}$ & Chemical formula / composition \\
\midrule
\endhead
\midrule
\insertTableNotes
\endlastfoot

Polyethylene & 330--370 & Yes; melts, drips and burns readily & 17--18 & Strongly enhanced burning; rapid flame spread and molten dripping are expected & 10--20 & 45 & \ce{(C2H4)_n} \\
Paper & 230--450 & Yes, if dry & 18--20 & Ignition and flame spread become easier; moisture content remains important & 10--15 & 14 & \ce{(C6H10O5)_n} \\
Food waste & No single value; dry residue approximately 250--400 & Poor when wet; combustible after drying & Variable & Moisture dominates; dried organic residue burns much more readily in oxygen-rich media & 25--50 & 3--5 & Variable organic compounds, mainly carbohydrates, proteins and lipids \\
Wood & 300--500 & Yes, after drying and pyrolysis & 20--22 & Strongly enhanced combustion; lower external heat input is needed after drying & 10--15 & 5--15 & Cellulose, hemicellulose and lignin; approximately \ce{(C6H10O5)_n} for cellulose \\
Glass & -- & No; non-combustible & n/a & Inert diluent and heat sink & 10--25 & 0 & Mainly \ce{SiO2} \\
Garden waste & 250--400 after drying & Poor when wet; yes after drying & 20--22 & Oxygen enrichment promotes combustion after moisture removal & 0--10 & 3--5 & Variable lignocellulosic biomass, mainly cellulose, hemicellulose and lignin \\
Vegetable oil & Flash point usually above 200; autoignition commonly above 300 & Yes, if heated, vaporised, or absorbed on porous solids & n.d. & High risk when dispersed, wicked into porous solids, or heated in oxygen-rich media & 2--5 & 38--40 & Triglycerides; approximate unsaturated fatty-acid fragment \ce{C18H32O2} \\
Protein & 200--350 decomposition / ignition range & Combustible when dry; tends to char and produce nitrogen-containing gases & n.d. & Oxygen enrichment promotes oxidation after drying and pyrolysis & 5--10 & 23--25 & Variable organic nitrogen compounds; amino acids may be represented by simplified empirical formulae \\
Fat & Flash point usually above 200; autoignition commonly above 300 & Yes, if heated, vaporised, or absorbed on porous solids & n.d. & High risk in porous waste because of wicking, heating and possible self-heating & 2--5 & 38--40 & Long-chain lipids and triglycerides; e.g., fatty-acid fragments such as \ce{C17H35COOH} \\
Rubber & 300--400 & Yes; smoke-intensive combustion & 18--25 & Risk increases strongly with oxygen concentration and pressure; behaviour depends on formulation and additives & 2--5 & 33 & Natural or synthetic elastomers; natural rubber approximately \ce{(C5H8)_n} \\
Fabric & 250--450 & Depends on fibre blend; cotton-rich fabrics burn readily, wool-rich fabrics may self-extinguish & 18--30 & Oxygen enrichment can overcome partial flame resistance and promote sustained flaming & 3--10 & 15--30 & Variable: polyester, cotton, nylon, wool and blends \\
Cotton & 350--420 & Yes; burns readily when dry & 18 & Strongly enhanced ignition and flame spread & 3--5 & 12 & Cellulose, \ce{(C6H10O5)_n} \\
Aluminium & Bulk material: --; fine powder may ignite & No for bulk metal; yes for fine powder under suitable conditions & n/a & Bulk aluminium is mostly an inert heat sink; fine particles may oxidise vigorously in oxygen-rich media & 1--5 & 31 & \ce{Al} \\
Ferrous materials & Bulk material: --; fine particles behave differently & No for bulk metal under ordinary MSW combustion conditions & n/a & Mostly inert heat sink; fine iron particles may oxidise more actively & 1--3 & 10--15 & \ce{Fe} \\
Diesel & Flash point above approximately 52; autoignition approximately 210--260 & Yes, when vapour generation is sufficient & n.d. & Oxygen enrichment greatly increases fire intensity and ignition risk & n/a & 43 & Mixture of hydrocarbons; simplified representation \ce{C12H26} \\
Petrol & Flash point below room temperature; autoignition approximately 280--470 & Yes; highly flammable vapour--air mixtures form easily & n.d. & Very high risk; oxygen enrichment increases flame speed and explosion severity & n/a & 44 & Mixture of hydrocarbons; simplified representation \ce{C8H18} \\
Charcoal & 350--450 & Yes; smouldering and glowing combustion are possible & n.d. & Oxygen enrichment strongly accelerates glowing combustion and oxidation rate & n/a & 30 & Mostly \ce{C} \\
Water sewage mud (80\% \ce{H2O}) & -- until substantial drying occurs & No when wet; organic fraction may burn after drying & n/a & Acts mainly as a heat sink; oxygen enrichment matters only after drying & -- & 2 & \ce{H2O} and organic matter \\
\bottomrule
\end{longtable}
\end{ThreePartTable}
\end{landscape}
\setlength{\tabcolsep}{6pt}

\begin{table}[htbp]
\centering
\caption{Comparison of reference combustion temperatures obtained using the simplified Python heat-balance model and Aspen Plus simulation at different flue gas recirculation (FGR) rates. The deviation is calculated relative to the temperature rise above room temperature, \(T_{\mathrm{room}}=\qty{25}{\degreeCelsius}\).}
\label{tab:S2_python_aspen}
\small
\begin{tabular}{cccccc}
\toprule
FGR rate &
\begin{tabular}{c}Python oxy\\ \(\qty{}{\degreeCelsius}\)\end{tabular} &
\begin{tabular}{c}Python air\\ \(\qty{}{\degreeCelsius}\)\end{tabular} &
\begin{tabular}{c}Aspen oxy\\ \(\qty{}{\degreeCelsius}\)\end{tabular} &
\begin{tabular}{c}Difference\\ \(\qty{}{\degreeCelsius}\)\end{tabular} &
Deviation \\
\midrule
0\%  & 3075 & 1539 & 2959 & \(+116\) & \(+4.0\%\) \\
30\% & 2286 & 1165 & 2314 & \(-28\)  & \(-1.2\%\) \\
50\% & 1744 & 907  & 1824 & \(-80\)  & \(-4.4\%\) \\
70\% & 1175 & 638  & 1264 & \(-89\)  & \(-7.2\%\) \\
\bottomrule
\end{tabular}

\vspace{0.5em}
\begin{flushleft}
\footnotesize
The difference is defined as
\[
\Delta T = T_{\mathrm{Python,oxy}} - T_{\mathrm{Aspen,oxy}}.
\]
The percentage deviation is calculated relative to the temperature rise above room temperature:
\[
\delta =
\frac{T_{\mathrm{Python,oxy}} - T_{\mathrm{Aspen,oxy}}}
{T_{\mathrm{Aspen,oxy}} - T_{\mathrm{room}}}
\times 100\%.
\]
Only the oxy-combustion results are compared with Aspen because the Aspen reference simulations were used for the oxy-fuel configuration.
\end{flushleft}
\end{table}

\clearpage

\begin{landscape}
\small
\setlength{\tabcolsep}{3pt}
\renewcommand{\arraystretch}{1.15}

\begin{longtable}{
p{0.035\linewidth}
p{0.145\linewidth}
p{0.070\linewidth}
p{0.075\linewidth}
p{0.160\linewidth}
p{0.145\linewidth}
p{0.300\linewidth}
}
\caption{Data points used for the ASU-scale figure: oxygen-production capacity, specific electricity demand, technology class, status, and source.}
\label{tab:S3_asu_scale_data}
\\

\toprule
No. &
Data point &
Capacity, t \(\mathrm{O_2}\) day\(^{-1}\) &
\(e_{\mathrm{O_2}}\), kWh kg\(^{-1}_{\mathrm{O_2}}\) &
Technology / method &
Status &
Source / source value \\
\midrule
\endfirsthead

\caption[]{Data points used for the ASU-scale figure, continued.}
\\
\toprule
No. &
Data point &
Capacity, t \(\mathrm{O_2}\) day\(^{-1}\) &
\(e_{\mathrm{O_2}}\), kWh kg\(^{-1}_{\mathrm{O_2}}\) &
Technology / method &
Status &
Source / source value \\
\midrule
\endhead

\midrule
\multicolumn{7}{r}{Continued on next page}
\\
\endfoot

\bottomrule
\endlastfoot

1 &
AirSep PSA high-pressure &
10 &
0.889 &
PSA, nominal 93\% \(\mathrm{O_2}\), higher product pressure &
Commercial / vendor-supported &
Chart/AirSep brochure; approximately 750 kWh per short ton of product for high-pressure PSA oxygen plants, converted here to kWh kg\(^{-1}_{\mathrm{O_2}}\) on a 93\% oxygen-product basis \parencite{ChartAirSep2006}. \\

2 &
AirSep PSA low-pressure &
20 &
0.474 &
PSA, nominal 93\% \(\mathrm{O_2}\), low-pressure design &
Commercial / vendor-supported &
Chart/AirSep brochure; approximately 400 kWh per short ton of product for low-pressure PSA oxygen plants, converted here to kWh kg\(^{-1}_{\mathrm{O_2}}\) on a 93\% oxygen-product basis \parencite{ChartAirSep2006}. \\

3 &
AirSep VPSA tonnage &
50 &
0.285 &
VPSA, nominal 93\% \(\mathrm{O_2}\), low pressure &
Commercial / vendor-supported &
Chart/AirSep brochure; VPSA tonnage systems, approximately 240 kWh per short ton of product, converted here to kWh kg\(^{-1}_{\mathrm{O_2}}\) on a 93\% oxygen-product basis \parencite{ChartAirSep2006}. \\

4 &
Air Liquide VSA, low-capacity end &
40 &
0.265 &
VSA oxygen generator, 90--93\% \(\mathrm{O_2}\) &
Commercial / vendor-supported &
Air Liquide standard plants brochure; VSA oxygen generator, 40--150 t \(\mathrm{O_2}\) day\(^{-1}\), 90--93\% \(\mathrm{O_2}\), 265 kWh t\(^{-1}_{\mathrm{O_2}}\) = 0.265 kWh kg\(^{-1}_{\mathrm{O_2}}\) \parencite{AirLiquideStandardPlants2017}. \\

5 &
Air Liquide VSA, high-capacity end &
150 &
0.265 &
VSA oxygen generator, 90--93\% \(\mathrm{O_2}\) &
Commercial / vendor-supported &
Same Air Liquide VSA range; high-capacity endpoint of the 40--150 t \(\mathrm{O_2}\) day\(^{-1}\) commercial range \parencite{AirLiquideStandardPlants2017}. \\

6 &
Air Liquide SIGMA, lower-energy end &
110 &
0.280 &
Standard cryogenic ASU, 96--99.6\% \(\mathrm{O_2}\) &
Commercial / vendor-supported &
Air Liquide standard plants brochure; SIGMA ASU, 110--350 t \(\mathrm{O_2}\) day\(^{-1}\), 96--99.6\% \(\mathrm{O_2}\), 280--460 kWh t\(^{-1}_{\mathrm{O_2}}\). The lower-energy endpoint is used here \parencite{AirLiquideStandardPlants2017}. \\

7 &
Air Liquide SIGMA, upper-energy end &
350 &
0.460 &
Standard cryogenic ASU, 96--99.6\% \(\mathrm{O_2}\) &
Commercial / vendor-supported &
Same Air Liquide SIGMA range; upper-energy endpoint of 280--460 kWh t\(^{-1}_{\mathrm{O_2}}\) \parencite{AirLiquideStandardPlants2017}. \\

8 &
Air Liquide Yango, lower-energy end &
330 &
0.400 &
Packaged cryogenic ASU, 99.5--99.8\% \(\mathrm{O_2}\) &
Commercial / vendor-supported &
Air Liquide standard plants brochure; Yango ASU, 330--770 t \(\mathrm{O_2}\) day\(^{-1}\), 99.5--99.8\% \(\mathrm{O_2}\), 400--600 kWh t\(^{-1}_{\mathrm{O_2}}\). The lower-energy endpoint is used here \parencite{AirLiquideStandardPlants2017}. \\

9 &
Air Liquide Yango, upper-energy end &
770 &
0.600 &
Packaged cryogenic ASU, 99.5--99.8\% \(\mathrm{O_2}\) &
Commercial / vendor-supported &
Same Air Liquide Yango range; upper-energy endpoint of 400--600 kWh t\(^{-1}_{\mathrm{O_2}}\) \parencite{AirLiquideStandardPlants2017}. \\

10 &
Experimental VPSA &
1 &
0.889 &
VPSA, 94\% \(\mathrm{O_2}\), semi-technical scale &
Experimental / semi-technical &
Energy consumption reported as 3200 MJ t\(^{-1}_{\mathrm{O_2}}\) for a semi-technical VPSA adsorption unit producing 94\% oxygen; converted here to 0.889 kWh kg\(^{-1}_{\mathrm{O_2}}\) \parencite{e20040232}. \\

11 &
PSA including compression &
30 &
0.680 &
PSA with product compression to 90 psig &
Literature / engineering case &
OSI Oxygen VPSA white paper; comparison value for PSA oxygen generation including compression to 90 psig, reported as 0.68 kWh kg\(^{-1}_{\mathrm{O_2}}\) \parencite{OSIOxygenVPSAWhitePaper2022}. \\

12 &
VSA including recompression &
80 &
0.440 &
VSA with product recompression to 90 psig &
Literature / engineering case &
OSI Oxygen VPSA white paper; comparison value for VSA oxygen generation including recompression to 90 psig, reported as 0.44 kWh kg\(^{-1}_{\mathrm{O_2}}\) \parencite{OSIOxygenVPSAWhitePaper2022}. \\

13 &
Optimized adsorption/VPSA, low estimate &
200 &
0.403 &
Large optimized adsorption/VPSA system, approximately 95\% \(\mathrm{O_2}\) &
Model / estimate &
Energy consumption for large optimized adsorption-based air separation reported as 1450 MJ t\(^{-1}_{\mathrm{O_2}}\); converted here to 0.403 kWh kg\(^{-1}_{\mathrm{O_2}}\) \parencite{e20040232}. \\

14 &
Optimized adsorption/VPSA, high estimate &
200 &
0.500 &
Large optimized adsorption/VPSA system, approximately 95\% \(\mathrm{O_2}\) &
Model / estimate &
Energy consumption for large optimized adsorption-based air separation reported as 1800 MJ t\(^{-1}_{\mathrm{O_2}}\); converted here to 0.500 kWh kg\(^{-1}_{\mathrm{O_2}}\) \parencite{e20040232}. \\

15 &
High-temperature membrane, vacuum process &
100 &
0.140 &
Prospective high-temperature membrane oxygen-separation process &
Prospective / theoretical &
Fraunhofer IKTS high-temperature oxygen membrane data; vacuum process electricity demand reported as approximately 0.2 kWh m\(^{-3}_{\mathrm{O_2}}\), converted here to 0.140 kWh kg\(^{-1}_{\mathrm{O_2}}\). This point is shown only as a prospective technology limit and is not used as a base-case assumption \parencite{FraunhoferIKTSOxygenMembrane}. \\

16 &
Cryogenic ASU-51 &
800 &
0.447 &
Liquid-pump cryogenic ASU &
Literature / design case &
Oxygen specific power consumption comparison for cryogenic ASUs; ASU-51 reported as 0.639 kWh Nm\(^{-3}_{\mathrm{O_2}}\), converted here to 0.447 kWh kg\(^{-1}_{\mathrm{O_2}}\) \parencite{Alsultanny2014OxygenSpecificPower}. \\

17 &
Cryogenic ASU-71 &
1000 &
0.442 &
Liquid-pump cryogenic ASU &
Literature / design case &
Oxygen specific power consumption comparison for cryogenic ASUs; ASU-71 reported as 0.631 kWh Nm\(^{-3}_{\mathrm{O_2}}\), converted here to 0.442 kWh kg\(^{-1}_{\mathrm{O_2}}\) \parencite{Alsultanny2014OxygenSpecificPower}. \\

18 &
Cryogenic ASU-31 &
1200 &
0.325 &
Low-pressure cryogenic ASU &
Literature / design case &
Oxygen specific power consumption comparison for cryogenic ASUs; ASU-31 reported as 0.464 kWh Nm\(^{-3}_{\mathrm{O_2}}\), converted here to 0.325 kWh kg\(^{-1}_{\mathrm{O_2}}\) \parencite{Alsultanny2014OxygenSpecificPower}. \\

19 &
Large-scale cryogenic ASU benchmark &
3000 &
0.196 &
Large-scale cryogenic ASU &
Literature / benchmark &
Large-scale cryogenic ASU benchmark reported as 0.28 kWh Nm\(^{-3}_{\mathrm{O_2}}\), converted here to 0.196 kWh kg\(^{-1}_{\mathrm{O_2}}\). This value supports the interpretation of 0.20 kWh kg\(^{-1}_{\mathrm{O_2}}\) as a favourable large-scale target \parencite{Alsultanny2014OxygenSpecificPower}. \\

20 &
Reference oxy-fuel MSW case &
197.5 &
0.265 &
ASU penalty assumed for this study &
Model assumption based on commercial VSA value &
Reference case calculated from the Aspen-based oxy-fuel MSW model: \(8231~\mathrm{kg\,h^{-1}}\times24~\mathrm{h}=197.5~\mathrm{t\,day^{-1}}\). The specific electricity demand was set to 0.265 kWh kg\(^{-1}_{\mathrm{O_2}}\), based on the Air Liquide VSA commercial value \parencite{AirLiquideStandardPlants2017}. \\

\end{longtable}

\vspace{0.5em}

\noindent\footnotesize
\textit{Notes.} Capacity is given as tonnes of oxygen per day. Specific electricity demand is expressed as kWh kg\(^{-1}_{\mathrm{O_2}}\). Vendor-supported values were taken from Air Liquide and Chart/AirSep brochures \parencite{AirLiquideStandardPlants2017,ChartAirSep2006}. Adsorption-based experimental and optimized values were taken from the adsorption energy-consumption review. PSA/VSA recompression values were taken from the OSI Oxygen VPSA white paper \parencite{OSIOxygenVPSAWhitePaper2022}. Cryogenic ASU design and benchmark values were taken from the oxygen specific power comparison study \parencite{Alsultanny2014OxygenSpecificPower}. The membrane point is included only as a prospective technology limit \parencite{FraunhoferIKTSOxygenMembrane}. The reference oxy-fuel MSW case is not a commercial ASU product, but a process-model point using the commercial VSA value \(e_{\mathrm{O_2}}=0.265~\mathrm{kWh\,kg^{-1}_{O_2}}\).

\end{landscape}

\subsection*{Gas-phase transport and condensate uptake of flue-gas impurities}

To support the assessment of condensation-assisted flue-gas cleaning, the
transport and uptake properties of the main gaseous impurities were considered.
Table~\ref{tab:S4_condensation_species} summarizes their estimated gas-phase
diffusion coefficients and representative literature data on interaction with
aqueous droplets.

The diffusion coefficients at \(150\,^{\circ}\mathrm{C}\) describe molecular
transport in the hot \(\mathrm{CO_2}\)-rich flue gas before condensation starts.
Additional values at \(70\,^{\circ}\mathrm{C}\) are given as representative
conditions inside the condensation-filtration zone. The ability of a species
to be removed by the condensate depends not only on gas-phase diffusion, but
also on accommodation at the gas--liquid interface, aqueous solubility, and
chemical reactions in the liquid phase.

\begin{landscape}

\footnotesize
\begin{longtable}{
L{1.6cm}
C{2.0cm}
C{2.0cm}
C{2.0cm}
L{4.2cm}
L{6.0cm}
L{3.4cm}
}

\caption{Gas-phase transport and aqueous uptake properties of selected flue-gas
impurities relevant to condensation-assisted exhaust cleaning.}
\label{tab:S4_condensation_species}
\\

\toprule
Species &
\(\boldsymbol{D_{150}}\),
\(10^{-5}\,\mathrm{m^2\,s^{-1}}\) &
\(\boldsymbol{D_{70}}\),
\(10^{-5}\,\mathrm{m^2\,s^{-1}}\) &
\(\boldsymbol{\tau_g}\) at \(70\,^{\circ}\mathrm{C}\), ms &
Representative accommodation / uptake parameter &
Main aqueous-phase behaviour &
Expected capture by condensate \\
\midrule
\endfirsthead

\caption[]{Gas-phase transport and aqueous uptake properties of selected
flue-gas impurities, continued.}
\\

\toprule
Species &
\(\boldsymbol{D_{150}}\),
\(10^{-5}\,\mathrm{m^2\,s^{-1}}\) &
\(\boldsymbol{D_{70}}\),
\(10^{-5}\,\mathrm{m^2\,s^{-1}}\) &
\(\boldsymbol{\tau_g}\) at \(70\,^{\circ}\mathrm{C}\), ms &
Representative accommodation / uptake parameter &
Main aqueous-phase behaviour &
Expected capture by condensate \\
\midrule
\endhead

\bottomrule
\endlastfoot

\ce{HCl} &
2.61 &
1.81 &
0.22 &
\(\alpha=0.13\)--0.24 at 263--281 K
\parencite{Schweitzer2000HCl} &
Rapid dissolution and dissociation to
\(\mathrm{H_3O^+}\) and \(\mathrm{Cl^-}\). &
High; depends on condensate acidity and renewal. \\

\ce{HF} &
3.47 &
2.41 &
0.17 &
No robust pure-water droplet value was used here. &
Strong interaction with water and aqueous fluoride chemistry;
partitioning depends strongly on condensate composition and acidity
\parencite{Cheng2018HF}. &
High in fresh or weakly acidic condensate; more composition-dependent
than HCl. \\

\ce{SO2} &
1.87 &
1.30 &
0.31 &
\(\alpha=0.175\pm0.015\) at 291 K
\parencite{Boniface2000SO2} &
Dissolution followed by formation of
\(\mathrm{HSO_3^-}\) and \(\mathrm{SO_3^{2-}}\);
uptake increases strongly with increasing pH. &
Moderate to high; strongly dependent on pH and liquid-phase chemistry. \\

\ce{SO3} &
1.80 &
1.25 &
0.32 &
Reactive uptake \(\gamma>0.7\) at approximately 300 K on concentrated
aqueous sulfuric acid
\parencite{Davidovits2011GasLiquid} &
Very rapid hydration and conversion to sulfuric acid after contact
with water. &
Very high once a liquid-water surface is available. \\

\ce{H2SO4} &
1.60 &
1.11 &
0.36 &
\(0.43<\alpha<1.0\) at 303 K; best-fit value approximately 0.65
\parencite{Poschl1998H2SO4} &
Very low volatility after uptake; remains in the condensed acidic
phase. &
Very high. \\

\ce{NO2} &
2.73 &
1.89 &
0.21 &
Overall aqueous uptake is typically of order
\(\gamma\sim10^{-4}\)--\(10^{-3}\) near 273--298 K
\parencite{JPL2003Evaluation14} &
Limited physical solubility; uptake is coupled to aqueous hydrolysis
and formation of nitrous and nitric acids
\parencite{Cheung2000NO2}. &
Low to moderate; reaction-limited and substantially less efficient
than acid-gas capture. \\

\ce{HNO3} &
2.23 &
1.55 &
0.26 &
Reported sticking probability approximately
\(0.3\)--1.0 for condensing \ce{HNO3}-\ce{H2O} mixtures
\parencite{Rudolf2001HNO3} &
Strong aqueous partitioning and dissociation to nitrate. &
High. \\

\ce{N2O} &
2.15 &
1.49 &
0.27 &
No reactive uptake coefficient is assumed. &
No rapid aqueous chemical sink comparable with HCl, \ce{SO2},
or \ce{HNO3}; physical dissolution alone is limited
\parencite{Sander2023Henry}. &
Low; condensation filtration should not be assumed to provide
efficient \ce{N2O} removal. \\

\end{longtable}

\vspace{0.5em}

\noindent\footnotesize
\textit{Notes.}
\(D_{150}\) and \(D_{70}\) are screening estimates of the binary gas-phase
diffusion coefficient of each trace species in \(\mathrm{CO_2}\) at a total
pressure of 1 bar. The Fuller--Schettler--Giddings correlation was used
\parencite{Fuller1966Diffusion}. Atomic and molecular diffusion-volume
increments were taken from the Fuller--Ensley--Giddings parameterization
\parencite{Fuller1969DiffusionVolumes}. The actual effective diffusivity in
multicomponent flue gas will differ somewhat from the binary values reported
here.

The characteristic gas-side transport time was estimated using the same
screening approximation as in the main text,
\[
\tau_g \sim \left(k_g a\right)^{-1},
\qquad
k_g\simeq\frac{D_i}{r},
\]
with \(r=2~\mu\mathrm{m}\) and \(a=500~\mathrm{m^{-1}}\). Thus,
\[
\tau_g \sim \frac{r}{D_i a}.
\]
The calculated values of approximately \(0.17\)--\(0.36\) ms indicate that
gas-phase molecular transport to the droplet surface is rapid under the
assumed conditions. Therefore, for soluble pollutants the overall removal
rate may instead be controlled by interfacial accommodation, aqueous
solubility, chemical reaction, droplet growth, and residence time.

Here, \(\alpha\) denotes the mass accommodation coefficient, whereas
\(\gamma\) denotes an overall or reactive uptake coefficient. These
quantities are not interchangeable. The literature values were measured
mainly near room temperature and are included only as representative
indicators of gas--liquid interaction. They should not be interpreted as
measured coefficients at \(70\,^{\circ}\mathrm{C}\) or
\(150\,^{\circ}\mathrm{C}\).

At \(150\,^{\circ}\mathrm{C}\), the reference oxy-fuel exhaust considered in
this study is above its water dew point, so a stable bulk liquid-water phase
is not expected. Consequently, \(D_{150}\) characterizes transport before
the condensation zone, while the accommodation and uptake parameters become
relevant only after liquid droplets or wetted particulate surfaces are
formed. \ce{SO3} and \ce{H2SO4} are listed separately as limiting chemical
forms; in wet flue gas their relative abundance depends on hydration,
temperature, and acid-dew-point chemistry.

The ``expected capture'' column is qualitative and should not be interpreted
as a predicted plant-scale removal efficiency. Henry-law data and aqueous
partitioning are strongly temperature- and composition-dependent
\parencite{Sander2023Henry}.

\end{landscape}
\subsection*{Preliminary material selection for the condensation section}

The condensation section is expected to operate in a particularly corrosive
environment because water condensation is accompanied by the absorption of
acidic species, especially \ce{HCl}, \ce{SO2}/\ce{SO3}, and potentially
\ce{HF}. Material selection therefore depends on both temperature and
condensate composition.

Table~\ref{tab:S5_materials} provides a preliminary screening of candidate
metallic materials using corrosion data for aerated \(1~\mathrm{wt.\%}\)
hydrochloric acid. This solution is not intended to reproduce the actual
multicomponent condensate, but is used as a common reference environment for
comparing candidate materials \parencite{AlleimaHCl}.

\begin{landscape}

\begin{table}[htbp]
\centering
\footnotesize

\caption{Preliminary corrosion screening of candidate materials for the
condensation section using aerated \(1~\mathrm{wt.\%}\) HCl as a reference
environment. The data are intended for comparative material screening and do
not represent the complete chemistry of the predicted flue-gas condensate
\parencite{AlleimaHCl}.}
\label{tab:S5_materials}

\begin{tabular}{
L{4.0cm}
C{2.0cm}
C{2.0cm}
C{2.0cm}
C{2.0cm}
L{7.0cm}
}

\toprule

Material &
\(20\,^{\circ}\mathrm{C}\) &
\(50\,^{\circ}\mathrm{C}\) &
\(60\,^{\circ}\mathrm{C}\) &
\(80\,^{\circ}\mathrm{C}\) &
Interpretation for the proposed condensation section \\

\midrule

Carbon steel &
2 &
2 &
2 &
2 &
Not suitable for direct contact with acidic condensate without a
corrosion-resistant lining. \\

316L stainless steel &
0p &
1p &
2 &
2 &
General corrosion may be limited at low temperature, but localized
corrosion remains possible. Unsuitable for the hotter condensation zone. \\

904L stainless steel &
0p &
0p &
1p &
1ps &
More resistant than 316L, but pitting and stress-corrosion risks become
important as temperature increases. \\

254 SMO &
0ND &
0ND &
0ND &
0ND &
Promising candidate for dilute acidic condensate. Localized-corrosion
resistance should nevertheless be confirmed using the actual condensate
composition. \\

654 SMO &
0ND &
0ND &
0ND &
0ND &
Highly resistant candidate for the acidic condensation zone; suitable as
one of the reference materials for experimental testing. \\

Super Duplex 2507 &
0ND &
0ND &
0ND &
0ND &
Potential candidate if the acidity and chloride concentration remain within
a confirmed acceptable range. \\

Commercially pure Ti &
0 &
0 &
0 &
0 &
Excellent resistance in the HCl reference environment. However, its use
cannot be recommended before the concentration of \ce{HF}/\ce{F^-} in the
condensate is established. \\

\bottomrule

\end{tabular}

\vspace{0.6em}

\begin{minipage}{0.96\linewidth}
\footnotesize
\textit{Notes.}
The corrosion-rating notation follows the source data:
0 = general corrosion rate \(<0.1~\mathrm{mm\,yr^{-1}}\);
1 = \(0.1\)--\(1~\mathrm{mm\,yr^{-1}}\);
2 = \(>1~\mathrm{mm\,yr^{-1}}\);
p = risk of pitting or crevice corrosion;
s = risk of stress-corrosion cracking;
ND = insufficient data for assessment of localized corrosion.
Therefore, ``0ND'' should not be interpreted as proof that localized
corrosion cannot occur.

The HCl solution is used here only as a screening reference. The actual
condensate may contain chlorides, fluorides, sulfates, dissolved
\ce{SO2}/\ce{SO3}, metal ions, ash particles, and other species that can
substantially change corrosion behaviour.
\end{minipage}

\end{table}

\end{landscape}

The screening results also show that material performance can change strongly
with acid concentration. At higher HCl concentrations, several alloys that
perform well in dilute acid become unsuitable at elevated temperature.
Therefore, Table~\ref{tab:S5_materials} should not be interpreted as a final
material specification. Final selection requires experimental testing using a
condensate composition representative of the proposed oxy-fuel MSW process.

For the most aggressive hot-condensation region, a Ni--Cr--Mo alloy such as
C-276 can be used as a conservative reference material. Manufacturer data
report corrosion rates of approximately \(0.01~\mathrm{mm\,yr^{-1}}\) in
\(2~\mathrm{wt.\%}\) HCl at \(66\,^{\circ}\mathrm{C}\) and
\(0.02~\mathrm{mm\,yr^{-1}}\) at \(79\,^{\circ}\mathrm{C}\)
\parencite{HaynesC276}. Titanium is also highly resistant to HCl, but fluoride
ions can destabilize its protective oxide film; therefore, Ti should not be
selected before the \ce{HF}/\ce{F^-} concentration has been quantified
\parencite{TiFluoride2021}.

\clearpage


\FloatBarrier

\section{Only-oxygen case}

The profit can be generated by realization of the product of incineration: electricity, heat, products of air separation (nitrogen, argon), solidified slag in the form of gravel or rock fiber and waiving of the cost associated with waste handling, namely landfill tax and gate fee of the landfill (transportation expenses are not included into consideration). Economic model depends on the type of air separation unit (ASU): industrial grade oxygen only (Linde LOGO) or complete separation of atmospheric air to oxygen, nitrogen and argon (Linde LION).

In the case of ``oxygen only'' ASU consumption of electricity is about \qtyrange{0.15}{0.2}{\kilo\watt\hour} per \qty{1}{\kilogram} of oxygen, while in the case of ASU providing complete separation of air requires about \qty{0.6}{\kilo\watt\hour} per \qty{1}{\kilogram} of oxygen, and ASU is more expensive than ``oxygen only'' ASU. Table ~\ref{tab:4} illustrates the amount of products and consumables associated with incineration of \qty{1}{\kilogram} of waste using a proposed scheme in the case of the UK, where electricity price and landfill tax are highest in Europe.

\FloatBarrier
\begin{table}\centering\footnotesize
 \caption{Estimated products (with ``$+$'' sign) and consumables (with ``$-$'' sign), its energy and money equivalent per \qty{1}{\kilogram} of waste \cite{ref:a1,ref:a2, ref:a3}}
 \label{tab:4}
\begin{tabular}{llllllll}
 \hline
 & \parbox[c][1.1cm]{1.4cm}{Required\\\ch{O2}} &
 \parbox[c][1.1cm]{1.5cm}{Produced\\\ch{CO2}} &
 \parbox[c][1.5cm]{1.5cm}{Produced\\electric\\energy} &
 \parbox[c][1.1cm]{1.5cm}{Produced\\gravel} & 
 \parbox[c][1.1cm]{1.3cm}{Landfill\\tax} &
 \parbox[c][1.1cm]{1.3cm}{Gate\\fee} &
 Net \\
 \hline
 Mass &
 \qty[print-implicit-plus=true]{-1.5}{\kilogram} &
 \qty[print-implicit-plus=true]{+2}{\kilogram} & &
 \qty[print-implicit-plus=true]{0.1}{\kilogram} &
 \qty[print-implicit-plus=true]{1}{\kilogram} &
 \qty[print-implicit-plus=true]{1}{\kilogram} & \\
 \parbox[c][1.1cm]{1.3cm}{Electric\\energy} &
 \qty[print-implicit-plus=true]{-0.3}{\kilo\watt} & &
 \qty[print-implicit-plus=true]{1}{\kilo\watt} & \\
 \parbox[c][1.1cm]{1.3cm}{Unit\\cost} &
 \qty{0.2}{\gbp\per\kilo\watt\per\hour} &
 \qty{0.2}{\gbp\per\kilogram} &
 \qty{0.2}{\gbp\per\kilo\watt\per\hour} &
 \qty{0.05}{\gbp\per\kilogram} &
 \qty{0.125}{\gbp\per\kilogram} &
 \qty{0.05}{\gbp\per\kilogram} \\
 Gain &
 \qty{-0.06}{\gbp} &
 \qty[print-implicit-plus=true]{+0.4}{\gbp}  &
 \qty[print-implicit-plus=true]{+0.14}{\gbp} &
 \qty[print-implicit-plus=true]{+0.005}{\gbp} &
 \qty[print-implicit-plus=true]{+0.125}{\gbp} &
 \qty[print-implicit-plus=true]{+0.05}{\gbp} &
 \parbox[c][1.1cm]{1.4cm}{\qty[print-implicit-plus=true]{+0.72}{\gbp}\\%
 \qty[print-implicit-plus=true]{+0.32}{\gbp}$^*$} \\
 \hline
 \end{tabular}
\end{table}

\begin{align*}
    &-\qty{1.5}{\kilogram} \times \qty{0.2}{\kilo\watt\per\kilogram} \times \qty{0.2}{\gbp\per\kilo\watt\per\hour} \,+
    & &\text{Electricity costs for \ch{O2} production} \\
    &\qty[print-implicit-plus=true]{+2}{\kilogram} \times \qty{0.2}{\gbp\per\kilogram} \,+
    & &\text{Produced \ch{CO2}} \\
    &\qty[print-implicit-plus=true]{+1}{\kilo\watt\hour} \times \qty{0.2}{\gbp\per\kilo\watt\per\hour} \,+
    & &\text{Produced electric energy}\\
    & \qty[print-implicit-plus=true]{+0.1}{\kilogram} \times \qty{0.005}{\gbp\per\kilogram}\, +
    & &\text{Produced gravel}\\
    & \qty[print-implicit-plus=true]{1}{\kilogram}\times \left( 0.125 + 0.05 \right) \unit{\gbp\per\kilogram} \,=
    & &\text{Landfill tax + gate fee}\\
    &= \qty[print-implicit-plus=true]{+0.72}{\gbp}
    & &\text{Total gain}\\
    \hline
    &= \qty[print-implicit-plus=true]{+0.32}{\gbp}
    & &\text{Total gain without \ch{CO2} sale}\\
    \hline
\end{align*}
CAPEX for ASU can be estimated as 70 000 GBP/ Ton-per-day \cite{ref:a4,ref:a5}. Note that such CAPEX is achieved when productivity exceeds \qty{400000}{\tonne\per\day}, for smaller rates it is slightly higher. For medium town with \num{100 000} habitant daily production of waste is about \qty{100}{\tonne\per\day} (which corresponds to approximately \qty{1}{\kilogram} of waste per second or heat power \qty{10}{\mega\watt}), which requires about \qty{150}{\tonne\per\day} of oxygen (CAPEX about £\num{15 000 000}) and a steam turbine unit with electric power \qty{3}{\mega\watt} (CAPEX about £\num{1 000 000}). Note that relative CAPEX for ASU achieves reduced constant.

Oven for incineration is based on standard engineering solutions used in metallurgy, steam generators are standard. Thus, CAPEX for incinerator, steam generator and filtering system could be estimated as £\num{5 000 000}. Cash flow (without considering possible \ch{CO2}) is about £\qty{32 000}{\perday} (£\qty{11 680 000}{\peryear})

Thus, the investments could be returned in about \qtyrange{2}{3}{\year}. In this case, the realization of \ch{CO2} and heat for industry cash flow could be doubled.
For larger scale systems, performance will be better, due to smaller relative OPEX and better technological efficiency. Finally, an estimate of the proposed scheme for London (with about \num{10 000 000} habitants). Approximately \qty{10 000}{\tonne} of waste are produced and about \qty{15000}{\tonne\per\day} of oxygen required. Total electric power of steam turbines is \qty{300}{\mega\watt} (around \qty{7}{\percent} of average consumption power of London (\qty{4}{\giga\watt})). The cost of ASU is of the order of £\num{1 000 000 000}. Cash flow (without considering possible realization of \ch{CO2} and heat) is about £\qty{3 200 000}{\perday} (£\qty{1 680 000 000}{\peryear}). Interest fee is £\qty{50 000 000}{\peryear} (assuming Bank of England rate \qty{5}{\percent}). Final economic performance will depend on the scheme of utilization of produced carbon dioxide and heat. Suppliers of modern high power steam turbines (Siemens) declare efficiency of high-performance steam turbines as high as \qty{45}{\percent}.

\section{Economic model based on complete air separation producing oxygen, nitrogen and argon}

In this case electricity amount required for separation of the atmospheric air is about \qty{0.6}{\kilo\watt\hour} per \qty{1}{\kilogram} of oxygen, but byproducts like nitrogen and argon are obtained. Large quantities of nitrogen, carbon dioxide heat energy produced make sensible production of fertilizer urea \ch{CO(NH2)2} on-site using chemical reaction:   
\ch{2 NH3 + CO2 -> CO(NH2)2 + H2O + 8 CO2} (\qty{150}{\degreeCelsius} \qty{250}{\bar}),
Ammonia \ch{NH3}  can be produced on-cite using the standard from Gaber-Bosh process: 
\ch{N2 + 3 H2 -> 2 NH3} (\qty{400}{\degreeCelsius} \qtyrange{150}{250}{\bar}),
while hydrogen can be obtained natural gas (methane \ch{CH4}) using steam process
\ch{CH4 + 2 H2O -> CO2 + 4 H2} (\qtyrange{700}{1000}{\degreeCelsius} \qtyrange{20}{30}{\bar}).

Using  the formulae one can estimate quantities of chemical substances associated with the incineration of \qty{1}{\kg} of waste and production of urea and corresponding economic effect which is illustrated in Table~\ref{tab:B2}. 

Such schemes require supply of natural gas and construction of ammonia production units.

In principle, hydrogen (and subsequently ammonia) can be produced on-cite using electrolysis of water, requiring \qty{1.7}{\kilo\watt} per \qty{1}{\kilogram} of waste. In this case will be required additional electricity from the grid.

\begin{table}\centering\footnotesize
 \caption{Estimated products (with ``$+$'' sign) and consumables (with ``$-$'' sign), related to the process of urea, money equivalent for \qty{1}{\kilogram} of waste \cite{ref:a6,ref:a7}}
 \label{tab:B2}
\begin{tabular}{lllll}
 \hline
 Compound &
 \parbox[c][1.1cm]{2.0cm}{Molar weight\\(\unit{\gram\per\mole})} &
 \parbox[c][1.1cm]{2.1cm}{Quantity per\\\qty{1}{\kilogram} of waste} &
 \parbox[c][1.1cm]{2.1cm}{Estimated\\price (\unit{\text{\$}\per\kilogram})} &
 \parbox[c][1.1cm]{1.9cm}{Realization\\price (\$)} \\
 \hline
 \ch{CO2}   & \num{44}    & \num[print-implicit-plus=true]{+2}    &   &   \\
 \ch{N2}    & \num{32}    & \num[print-implicit-plus=true]{+6}    &   &   \\
 \ch{Ar}    & \num{40}    & \num[print-implicit-plus=true]{+0.08} & \num{0.5}   & \num[print-implicit-plus=true]{+0.04} \\
 \ch{NH3}   & \num{17}    & \num[print-implicit-plus=true]{-1.5}  &   &   \\
 \ch{H2}    & \num{2}     & \num[print-implicit-plus=true]{-0.17} &   &   \\
 \ch{CH4}   & \num{16}    & \num[print-implicit-plus=true]{-0.5}  &   &   \\
 \ch{CO(NH2)2}  &   & \num[print-implicit-plus=true]{+2.7}  & \num{0.35}  & \num[print-implicit-plus=true]{+0.95} \\
 \hline
 \end{tabular}
\end{table}







\clearpage

\clearpage
\section*{Supplementary references}

\end{document}